\documentclass[amsmath,amssymb,
showpacs,twocolumn,
superscriptaddress,
prl]{revtex4-1}
\usepackage{graphicx,bm,color,subfigure}
\usepackage[T1]{fontenc}
\usepackage{mathrsfs}
\usepackage{bm}
\usepackage{amsmath}
\usepackage{amssymb}
\usepackage{graphicx}
\usepackage{esint}
\usepackage{multirow}
\usepackage{float}
\usepackage{array}
\usepackage{makecell}
\usepackage{harpoon}
\usepackage{booktabs}
\usepackage{gensymb}
\usepackage{simplewick}
\usepackage{subfigure}
\usepackage{soul}


\begin{document}

\title{Charge-4e superconductivity and chiral metal in the 45\degree-twisted bilayer cuprates and similar materials}

\author{Yu-Bo Liu}
\thanks{These two authors contributed equally to this work.}
\affiliation{School of Physics, Beijing Institute of Technology, Beijing 100081, China}
\author{Jing Zhou}
\thanks{These two authors contributed equally to this work.}
\affiliation{Department of Science, Chongqing University of Posts and Telecommunications,
	Chongqing 400065, China}
\author{Congjun Wu}
\affiliation{Institute for Theoretical Sciences, WestLake University, Hangzhou 310024, China}
\affiliation{Department of Physics, School of Science, Westlake University, Hangzhou 310024, Zhejiang, China}
\affiliation{
Key Laboratory for Quantum Materials of Zhejiang Province, School of Science, Westlake University, Hangzhou 310024, China}
\affiliation{
Institute of Natural Sciences, Westlake Institute for Advanced Study, Hangzhou 310024, Zhejiang, China}
\author{Fan Yang}
\email{yangfan_blg@bit.edu.cn}
\affiliation{School of Physics, Beijing Institute of Technology, Beijing 100081, China}

\begin{abstract}
The material realization of the charge-4e/6e superconductivity (SC) is a big challenge. Here we propose realization of the charge-4e SC and chiral metal through stacking a homo-bilayer with the largest twist angle, forming the twist-bilayer quasi-crystal (TB-QC), exampled by the 45\degree-twisted bilayer cuprates and 30\degree-twisted bilayer graphene. When each mononlayer hosts a pairing state with the largest pairing angular momentum, previous studies yield that the second-order interlayer Josephson coupling would drive chiral topological SC (TSC) in the TB-QC. Here we propose that, above the $T_c$ of the chiral TSC, either the total- or relative- pairing phase of the two layers can be unilateral quasi-ordered or ordered, leading to the charge-4e SC or the chiral metal phase. Based on a thorough symmetry analysis to get the low-energy effective Hamiltonian, we conduct a combined renormalization-group and Monte-Carlo study and obtain the phase diagram, which includes the charge-4e SC and chiral metal phases.
\end{abstract}

\maketitle

The charge-4e/6e superconductivities (SCs) are exotic SCs characterized by $\frac{1}{2}$/$\frac{1}{3}$ flux quantization. These novel SCs are formed by condensation of electron quartets/sextets\cite{Korshunov1985,Kivelson1990,Ropke1998,Doucot2002,Babaev2004,DHLee2004,Wu2005, Aligia2005,Agterberg2008, Berg2009, Agterberg2011,Wen2009,Herland2010, You2012, Jiang2017, Zeng2021, Fernandes2021, Jian2021, Song2022,JPHu2022,Ge_J2022,Zhou2022,Hu_Xiao2022,Lee2022,Yuyue2022}, which is beyond the conventional Bardeen-Cooper-Schrieffer mechanism\cite{BCS_theory}. Recently, it was proposed that these intriguing SCs can emerge as the high-temperature vestigial phases of the charge-2e SC in systems hosting multiple coexisting pairing order parameters (ODPs). Typical proposals for such multi-component pairings include the incommensurate pair-density-wave (PDW)\cite{Berg2009, Agterberg2011,You2012}, the nematic pairing\cite{Fernandes2021, Jian2021} and the bilayer pairing system\cite{Song2022,Zeng2021}. However, each proposal is still waiting for the experiment realization.

One proposal is through melting of incommensurate PDW\cite{Berg2009, Agterberg2011,You2012}. The PDW has been reported in such materials as the cuprates~\cite{Bi_PDW,Bi_PDW1}, the CsV$_3$Sb$_5$~\cite{CsV3Sb5_PDW} and the transition-metal dichalcogenide~\cite{S_PDW}. This proposal, however, suffers from the difficulty that the PDWs observed in these experiments are always accompanied by a dominant uniform SC part.  Another proposal is through melting of nematic pairing\cite{Fernandes2021, Jian2021}. Such pairing state is formed through real mixing of the two basis functions of a two-dimensional (2D) irreducible representation (IRRP) of the point group. More recently, a group-theory based classification of the vestigial phases generated by melting of the pairing states belonging to the 2D IRRPs was performed~\cite{Hecker2303}, wherein such interesting phase as $d$-wave charge-4e SC was proposed. However, the experiment verification of these proposals are still on the way. Alternatively, a bilayer approach was recently proposed\cite{Song2022,Zeng2021} in which, two monolayers hosting SCs with different phase stiffness are coupled. Consequently, in an intermediate-temperature vestigial phase, one layer carries charge-2e SC while the other layer carries charge-4e SC\cite{Song2022}. The draw back of this proposal lies in that, in an out-of-plane magnetic field, while the charge-4e-SC layer allows for integer times of half magnetic flux, the charge-2e-SC layer only allows for integer flux. As the two layers experience the same magnetic flux, only the integer flux is allowed, and the hallmark of the charge-4e SC, i.e. the half flux quantization, cannot be experimentally detected in this proposal. Finally, the melting of the multi-component hexatic chiral superconductor leading to vestigial charge-6e SC was proposed in the context of kagome superconductors\cite{Zhou2022}.Presently, the material realization of the charge-4e/6e SC is still a big challenge.


Here in this work, we take advantage of the rapid development of the ``{\it twistronics}'' \cite{caoyuan20181,caoyuan20182,Dean2018,Chenguorui20191,P_Kim2020,Park2021,Regan2020,Tang2020,Yankowitz2019,Yazdani2019,Efetov2019,David2019,Serlin2019,Zeldov2020,Caoyuan2021,
Xu2018,Po2018,YangFan2018,WuFeng20181,Kang2018,Isobe2018,Koshino2018,Fernandes2018,Gonzalez2019,Song2019,Bultinck2020prx,Senthil2020,Chen_Lu2022,Valagiannopoulos2022}, and utilize it to design the intriguing charge-4e SC. Here we shall study materials made through stacking two identical monolayers with the largest twist angle, which host Moireless quasi-crystal (QC) structures\cite{Moon2019,Park2019,Yuan2020} and are dubbed as the twist-bilayer QC (TB-QC)\cite{Yu_Bo2023}, exampled by the recently synthesized 30\degree-twisted bilayer graphene\cite{Ahn2018,Yao2018,Pezzini2020,Yan2019,Deng2020} and 45\degree-twisted bilayer cuprates\cite{Zhu2021, Zhao2021}. Prominently, the TB-QC hosts a doubly-enlarged fold of rotation axis relative to its monolayer. Previous study\cite{Yu_Bo_new, Yu_Bo2023} suggests that when each monolayer hosts a pairing state carrying the largest pairing angular momentum for the lattice, the second-order interlayer Josephson coupling (IJC) between the pairing ODPs from the two layers in the TB-QC makes them to mix as $1:\pm i$, leading to time-reversal symmetry (TRS) breaking chiral topological SC (TSC). For example, as the monolayer cuprate carries the $d$-wave pairing, the 45\degree-twisted bilayer cuprates will host the $d+id$ chiral TSC\cite{cuprates_QC, JPHu2018, cuprates_QC2,cuprates_QC3,Yu_Bo2023}. It's interesting to investigate possible vestigial secondary orders above the $T_c$ of these chiral TSC phases, driven by the second-order IJC between the pairing ODPs from the two layers.

\begin{figure}[htbp]
\centering
\includegraphics[width=0.45\textwidth]{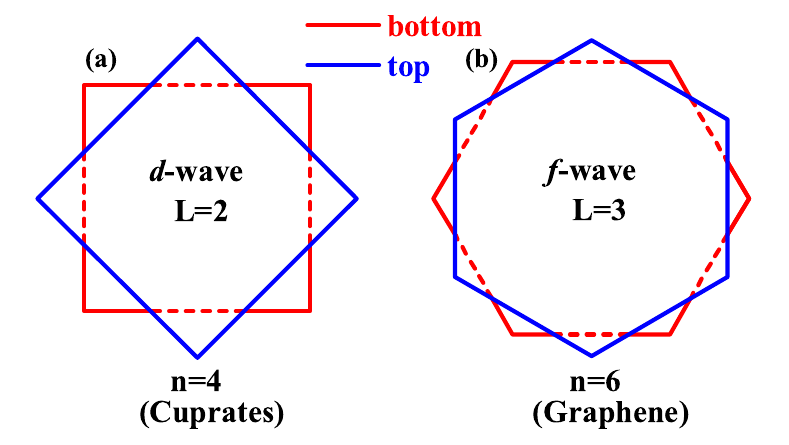}
\caption{(Color online) Schematic illustration of a TB-QC formed by two $D_n$-symmetric monolayers, with each monolayer carrying SC with pairing angular momentum $L=\frac{n}{2}$. We take $n=4$ (cuprates) and $n=6$ (graphene) for example. }\label{geometry}
\end{figure}

In this paper, we study the secondary orders in the superconducting TB-QC. Its unique symmetry leads to a simplified low-energy effective Hamiltonian including decoupled total- and relative- phase fields between the bilayer. Significantly, the second-order IJC allows the relative phase to fluctuate between its two saddle points to restore the TRS. Consequently, while the unilateral order of the relative-phase field leads to the TRS-breaking chiral-metal phase, the unilateral quasi- order of the total-phase field leads to the charge-4e SC phase, in which two Cooper pairs from different layers pair to form a quartets. These two vestigial phases occupy different regimes in the phase diagram obtained by our combined renormalization group (RG) and Monte-Carlo (MC) studies, which are unambiguously identified by various temperature-dependent quantities including the specific heat, the secondary ODPs and their susceptibilities, as well as the spatial-dependent correlation functions.

~~~~

\noindent{{\bf Results}}

~~~~~~

{\bf Model and Symmetry:} Taking two $D_n$-symmetric monolayers, let's stack them by the twist angle $\pi/n$ to form a TB-QC, as shown in Fig.~\ref{geometry} for $n=4$ (e.g. the cuprates) and $n=6$ (e.g. the graphene). Obviously, the point group is $D_{nd}$, isomorphic to $D_{2n}$. There is an additional symmetry generator in the TB-QC which is absent in its monolayer, i.e. the $C^{1}_{2n}$ rotation accompanied by a succeeding layer exchange, renamed as $\tilde{C}^{1}_{2n}$ here.

Suppose that driven by some pairing mechanism, the monolayer $\mu=\text{t/b}$ (top/bottom) can host a pairing state with pairing angular momentum $L=n/2$. While the cuprate monolayer hosting the $d$-wave SC synthesized recently\cite{Yuanbo2019} provides a good example for $n=4$, some members in the graphene family which were predicted to host the $f$-wave SC \cite{Kiesel2012,Yu_Bo_new, Benjamin2022} set an example for $n=6$. The pairing gap function in the $\mu$ layer is
\begin{equation}\label{gap_function}
\Delta^{(\mu)}(\mathbf{k})=\psi_{\mu}\Gamma^{(\mu)}(\mathbf{k}).
\end{equation}
Here $\Gamma^{(\mu)}(\mathbf{k})$ is the normalized real form factor, and $\psi_{\mu}$ is the ``complex pairing amplitude''. Prominently, the $\Gamma^{(\mu)}(\mathbf{k})$ for $L=n/2$ changes sign with every $C_n^1$ rotation. As shown in Fig.~\ref{geometry}, we choose a gauge so that
\begin{equation}\label{relation_gap_updn}
\Gamma^{(\text{b})}(\mathbf{k})=\hat{P}_{\frac{\pi}{n}}\Gamma^{(\text{t})}(\mathbf{k}),~~ \hat{P}_{\frac{2\pi}{n}}\Gamma^{(\mu)}(\mathbf{k})=-\Gamma^{(\mu)}(\mathbf{k}).
\end{equation}
Here $\hat{P}_{\phi}$ indicates the rotation by the angle $\phi$. As the interlayer coupling in the TB-QC is weak\cite{Moon2019, Park2019, Yu_Bo2023}, we can only consider the dominant intralayer pairing, but the two intralayer pairing ODPs can couple through the IJC\cite{Yu_Bo_new,cuprates_QC, JPHu2018, cuprates_QC2,cuprates_QC3,Benjamin2022}. We shall investigate the ground state and the vestigial secondary orders induced by this IJC.

Firstly, let's make a saddle-point analysis for the Ginzburg-Landau (G-L) free energy $F$ as functional of $\psi_{\text{t/b}}$. For the saddle-point solution, the $\psi_{\text{t/b}}$ are spatially uniform constant numbers. $F$ is decomposed as,
\begin{eqnarray}\label{G_L_F}
F\left(\psi_{\text{t}},\psi_{\text{b}}\right)&=&F_0(\left|\psi_{\text{t}}\right|^2)+F_0(\left|\psi_{\text{b}}\right|^2)+F_{J}\left(\psi_{\text{t}},\psi_{\text{b}}\right),
\end{eqnarray}
where $F_0(\left|\psi_{\mu}\right|^2)$ are the monolayers terms and $F_J$ is the IJC. The TRS-allowed first-order IJC takes the form,
\begin{eqnarray}\label{Josephson}
F^{(1)}_{J}\left(\psi_{\text{t}},\psi_{\text{b}}\right)&=&-\alpha\left(\psi_{\text{t}}\psi_{\text{b}}^*+c.c\right).
\end{eqnarray}
Under $\tilde{C}^{1}_{2n}$, the gap function on the $\mu$ layer changes from $\Delta^{(\mu)}(\mathbf{k})=\psi_\mu\Gamma^{\left(\mu\right)}(\mathbf{k})$ to $\tilde{\Delta}^{(\mu)}(\mathbf{k})=\psi_{\bar{\mu}}\hat{P}_{\frac{\pi}{n}}\Gamma^{\left(\bar\mu\right)}(\mathbf{k})$ which, under Eq. (\ref{relation_gap_updn}), can be rewritten as $\tilde{\psi}_\mu\Gamma^{\left(\mu\right)}(\mathbf{k})$ with
\begin{equation}\label{ODP_change}
\tilde{\psi_\text{b}}=\psi_\text{t}, ~~~~~~~~~~\tilde{\psi_\text{t}}=-\psi_\text{b}.
\end{equation}
The invariance of $F$ under $\tilde{C}^{1}_{2n}$ requires $\alpha=0$. Thus, the following second-order IJC should be considered,
\begin{eqnarray}\label{G_L_F_2}
F_J\left(\psi_{\text{t}},\psi_{\text{b}}\right)=A_0\left(\psi_{\text{t}}^2\psi_{\text{b}}^{2*}+{\rm c.c.}\right)+O\left(\psi^6\right).
\end{eqnarray}
Eq. (\ref{G_L_F_2}) is minimized at $\psi_b=\pm i \psi_t$ for $A_0>0$ or $\psi_b=\pm \psi_t$ for $A_0<0$. Previous microscopic calculations favor the former for the 45\degree-twisted bilayer cuprates\cite{Yu_Bo2023,cuprates_QC} and 30\degree-twisted bilayer of the graphene family\cite{Yu_Bo_new, Benjamin2022}, leading to $d+id$ or $f+if$ chiral TSCs ground state.

\begin{figure}[htbp]
	\centering
	\includegraphics[width=0.45\textwidth]{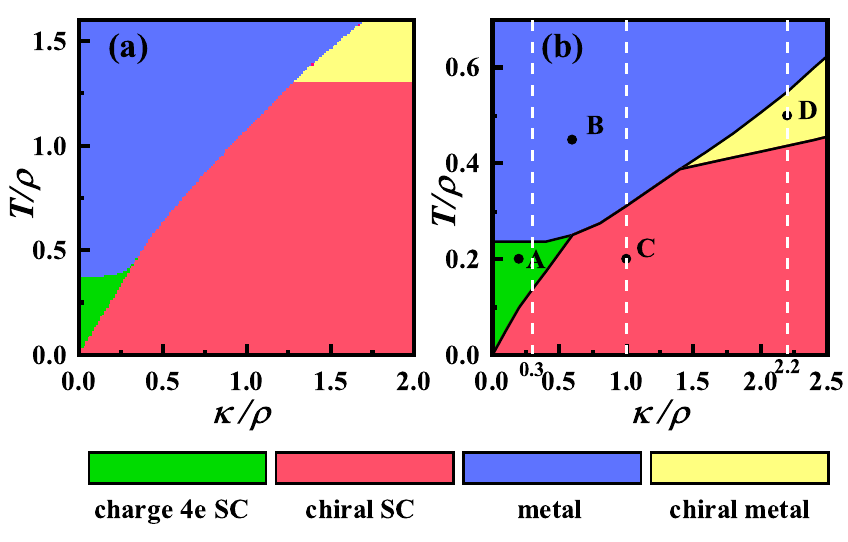}
	\caption{(Color online) Phase diagram provided by (a) the RG study and (b) the MC study. The initial values of the coupling parameters in (a) are $g_{2,0}=g_{0,2}=0.1$, $g_{1,1}=g_{4}=0.01$  in Eq. (\ref{eqn:action-SG}) and in (b) are $A=0.025\rho$ and $\gamma=\frac{1}{4}\rho\kappa/(\rho+\kappa)$  in Eq. (\ref{Hamiltonian_d}). }\label{phase_diagram}
\end{figure}

Secondly, let's provide the low-energy effective Hamiltonian for the pairing-phase fluctuations. In this study, we fix $\Gamma^{(\mu)}$ and set $\psi_\mu \to \psi_\mu\left(\mathbf{r}\right)$ as a slowly-varying ``envelope'' function to describe the spatial fluctuation of the complex pairing amplitude. Focusing on the phase fluctuation, $\psi_{\text{t/b}}$ are written as $\psi_{\text{t/b}}=\psi_{0}e^{i\theta_{\text{t/b}}\left(\mathbf{r}\right)}$ where $\psi_{0}>0$ is a constant. The $\theta_{\text{t/b}}\left(\mathbf{r}\right)$ are further written as
\begin{eqnarray}\label{phase_re_tl}
\theta_{\text{t}}\left(\mathbf{r}\right)=\theta_{+}\left(\mathbf{r}\right)+\theta_{-}\left(\mathbf{r}\right),~~~\theta_{\text{b}}\left(\mathbf{r}\right)=\theta_{+}\left(\mathbf{r}\right)-\theta_{-}\left(\mathbf{r}\right).
\end{eqnarray}
Here $\theta_{+}\left(\mathbf{r}\right)$ and $\theta_{-}\left(\mathbf{r}\right)$ denote the total and relative pairing phases. The low-energy effective Hamiltonian reads
\begin{eqnarray}\label{Hamiltonian_re}
H=H_0\left[\partial_{\pm}\theta_{+},\partial_{\pm}\theta_{-} \right]+A_0\psi_0^4\int \cos 4\theta_{-}\left(\mathbf{r}\right) d^2\mathbf{r},~~
\end{eqnarray}
with $\partial_{\pm}\equiv\partial_{x}\pm i\partial_{y}$. Up to the lowest-order expansion, the $H_0$ takes the following explicit form in the $\mathbf{k}$-space,
\begin{eqnarray}\label{Hamiltonian_k}
H_0&=&\frac{1}{2}\int d^{2}\mathbf{k}\left[\theta_{+}\left(\mathbf{k}\right)\theta_{+}\left(\mathbf{-k}\right)\left(\alpha k^2_{+}+\beta k^2_{-}+\rho k_{+}k_{-} \right)\right.\nonumber\\
&&~~+\theta_{+}\left(\mathbf{k}\right)\theta_{-}\left(\mathbf{-k}\right)\left(\omega k^2_{+}+\delta k^2_{-}+\eta k_{+}k_{-} \right)\nonumber\\
&&~~+\left.\theta_{-}\left(\mathbf{k}\right)\theta_{-}\left(\mathbf{-k}\right)\left(\epsilon k^2_{+}+\xi k^2_{-}+\kappa k_{+}k_{-} \right)\right].
\end{eqnarray}

Under $\tilde{C}_{2n}^1$, the gap function on the $\mu$ layer changes from $\Delta^{(\mu)}=\psi_\mu\Gamma^{\left(\mu\right)}$ to $\tilde{\Delta}^{(\mu)}=\psi_{\bar{\mu}}\hat{P}_{\frac{\pi}{n}}\Gamma^{\left(\bar\mu\right)}$ which, under Eq. (\ref{relation_gap_updn}), can be rewritten as $\tilde{\psi}_\mu\Gamma^{\left(\mu\right)}$ with
\begin{equation}
\tilde{\psi}_{\text{b}}\left(\mathbf{r}\right)=\psi_{\text{t}}\left(\hat{P}^{-1}_{\frac{\pi}{n}}\mathbf{r}\right),~~~ \tilde{\psi}_{\text{t}}\left(\mathbf{r}\right)=-\psi_{\text{b}}\left(\hat{P}^{-1}_{\frac{\pi}{n}}\mathbf{r}\right).
\end{equation}
Consequently, we have
\begin{equation}\label{phasek_change}
\theta_{+}\left(\mathbf{k}\right)\to\theta_{+}\left(\hat{P}^{-1}_{\frac{\pi}{n}}\mathbf{k}\right),~\theta_{-}\left(\mathbf{k}\right)\to-\theta_{-}\left(\hat{P}^{-1}_{\frac{\pi}{n}}\mathbf{k}\right).
\end{equation}
The invariance of Eq. (\ref{Hamiltonian_k}) under (\ref{phasek_change}) only allows for nonzero $\rho$ and $\kappa$, leading to the real-space Hamiltonian
\begin{eqnarray}\label{Hamiltonian_r}
H=\int d^{2}\mathbf{r}\left(\frac{\rho}{2}\left|\triangledown\theta_{+}\right|^2+ \frac{\kappa}{2}\left|\triangledown\theta_{-}\right|^2+A_0\psi_0^4 \cos 4\theta_{-}\right).~~~~~
\end{eqnarray}

Eq. (\ref{Hamiltonian_r}) shows two important features. Firstly, the $\theta_{+}$ and $\theta_{-}$ fields are dynamically decoupled, with each hosting different stiffness parameter $\rho$ or $\kappa$ derived by the G-L expansion in the Supplementary Material (SM)\cite{SM}. Secondly, the second-order IJC allows $\theta_\text{t}-\theta_{\text{b}}=2\theta_{-}$ to fluctuate between its two saddle points, i.e. $\pm\pi/2$, to restore the TRS. Note that although the term $\cos(4\theta_-)$ in Eq. (\ref{Hamiltonian_r}) leads to four different values of $\theta_-:  \pm\pi/4$ and $\pm3\pi/4$ for the ground state, $\theta_-=\pi/4$ ($-\pi/4$) leads to gauge equivalent state with $\theta_-=-3\pi/4$ ($3\pi/4$). So the system only possesses two-fold Ising anisotropy. Here the unilateral quasi-ordering of the $\theta_{+}$ field leads to the ODP $\Delta^{(\text{t})}(\mathbf{k})\cdot\Delta^{(\text{b})}(-\mathbf{k})$ characterizing the charge-4e SC in which two Cooper pairs from different layers pair. The unilateral ordering of the $\theta_{-}$ field leads to the ODP $\Delta^{(\text{t})*}(\mathbf{k})\cdot\Delta^{(\text{b})}(\mathbf{k})$ characterizing the TRS breaking chiral metal\cite{Asle2013,Grinenko2021}. Note that while $\theta_{+}$ and $\theta_{-}$ each can host either integer or half-integer vortices, Eq. (\ref{phase_re_tl}) requires that they can only simultaneously host integer or half-integer vortices to ensure the single-valuedness of $\psi_{\text{t/b}}$\cite{Berg2009,Jian2021}. This sets the ``kinematic constraint'' in the low-energy ``classical Hilbert space'' for allowed vortices of the two fields.

{\bf RG Study:} To perform the RG study, we start with the following effective action at the temperature $T$,
\begin{equation}
S=\int d^{2}\mathbf{r}\left( \frac{\rho}{2T}\left|\triangledown\theta_{+}\right|^2+\frac{\kappa}{2T}\left|\triangledown\theta_{-}\right|^2+g_{4}\cos4\theta_{-} \right) \label{eqn:action}
\end{equation}
Here $g_{4}>0$ is proportional to $A_0$. This action can be mapped to a two-component Sine-Gordon model,
\begin{eqnarray}
\label{eqn:action-SG}
S_{\mathrm{SG}}&&=\int d^{2}\mathbf{x}\left( \frac{T}{2\rho}\left|\triangledown\tilde{\theta}_{+}\right|^2+\frac{T}{2\kappa}\left|\triangledown\tilde{\theta}_{-}\right|^2+g_{4}\cos4\theta_{-} -g_{2,0}\right.\nonumber\\
&&\left.\times\cos2\pi\tilde{\theta}_{+}-g_{0,2}\cos2\pi\tilde{\theta}_{-}-g_{1,1}\cos\pi\tilde{\theta}_{+}\cos\pi\tilde{\theta}_{-}\right)~~~~~
\end{eqnarray}
The dual bosonic fields $\tilde{\theta}_{+}$ and $\tilde{\theta}_{-}$ describe the vortices of the fields $\theta_{+}$ and $\theta_{-}$. $g_{2,0}$, $g_{0,2}$ and $g_{1,1}$ are coupling parameters proportional to the fugacities of different types of vortices ($g_{2,0}/g_{0,2}$: integer vortices; $g_{1,1}$: half vortices).

The phase diagram obtained by the one-loop RG analysis provided in {\bf Method} is shown in Fig.~\ref{phase_diagram}(a). Variation of the initial coupling parameters doesn't change the topology of the phase diagram, which always include the chiral TSC, charge-4e SC, chiral metal and normal metal phases, see the SM~\cite{SM}. At low enough $T$, the vortex fugacities $g_{2,0}$, $g_{0,2}$ and $g_{1,1}$ are all irrelevant while the IJC parameter $g_{4}$ is relevant, suggesting that both the $\theta_{\pm}$ fields are locked, leading to the TRS breaking chiral SC. With the enhancement of $T$, in the low $\kappa/\rho$ regime, $g_{0,2}$ first gets relevant (and suppresses $g_4$ ) suggesting that the $\theta_{-}$ vortices proliferate to restore the TRS, to form the charge-4e SC. In the high $\kappa/\rho$ regime instead, the $g_{2,0}$ first gets relevant suggesting the $\theta_{+}$ vortices proliferate to kill the SC, to form the chiral metal. In both regimes, at high enough $T$, $g_{2,0}$ and $g_{0,2}$ are both relevant, forming the normal metal phase. In the regime $\kappa\approx\rho$, with the enhancement of $T$, the system transits into a phase wherein the coupling $g_{1,1}$ is relevant and the half vortices involving both fields proliferate to kill both (quasi) orders, suggesting that the system directly transit to the normal state.

In the charge-4e SC, the Josephson-coupling phase, i.e. $\theta_-$, is disordered. However, this phase should not be understood as a layer-decoupled charge-2e SC from each layer, as in this phase the pairing phase of each layer is also disordered. To remind, the charge-4e SC proposed here only lives in the intermediate temperature above the $T_c$ of the pairing state, wherein each layer is no longer superconducting. In the chiral metal phase, the time-reversal symmetry breaking can be verified by the polar Kerr effect. Further more, there can be spontaneously generated inner magnetic field in the material, which can be detected by the muon spin resonance experiment.

\begin{figure}[htbp]
	\centering
	\includegraphics[width=0.48\textwidth]{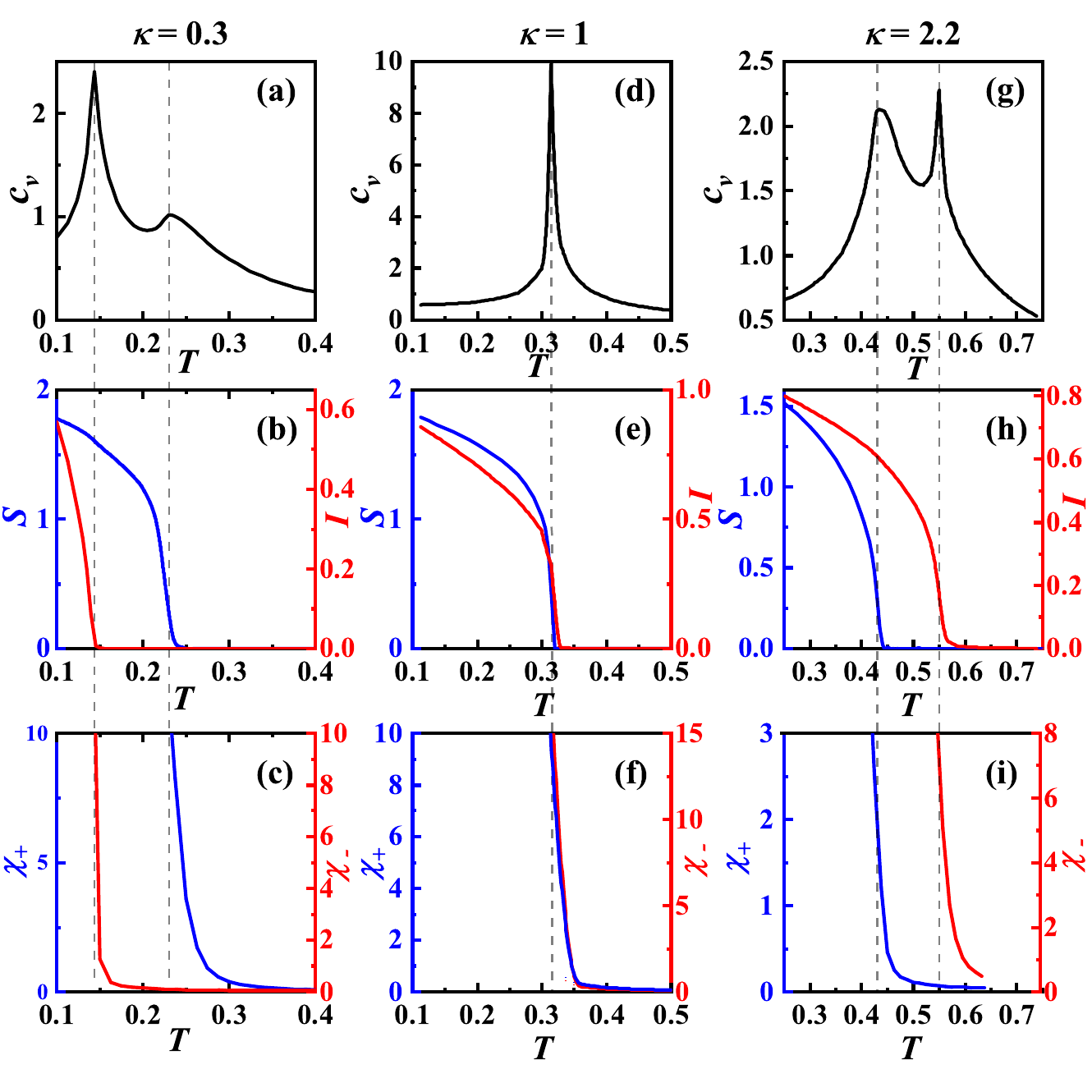}
	\caption{(Color online) Various $T$-dependent quantities for $\kappa=0.3$ (a-c), $\kappa=1$ (d-f) and $\kappa=2.2$ (g-i). (a), (d) and (g) The specific heat $C_v$. (b), (e) and (h) The phase stiffness $S$ (blue) and Ising ODP $I$ (red). (c), (f) and (i) The susceptibilities $\chi_{+}$ (bule) and $\chi_{-}$ (red). The $\rho$ is set as the unit of $\kappa$ and $T$.}\label{temperature_dependence}
\end{figure}

{\bf MC study:} To perform the MC study, we discretize the Hamiltonian (\ref{Hamiltonian_r}) on the square lattice to obtain
\begin{eqnarray}\label{Hamiltonian_d}
H &=&-\alpha\sum_{\langle ij\rangle}\cos[\theta_{\text{t}}(\mathbf{r}_i)+\theta_{\text{b}}(\mathbf{r}_i)-\theta_{\text{t}}(\mathbf{r}_j)-\theta_{\text{b}}(\mathbf{r}_j)]	 \nonumber\\
&-&\lambda\sum_{\langle ij\rangle}\cos[\theta_{\text{t}}(\mathbf{r}_i)-\theta_{\text{b}}(\mathbf{r}_i)-\theta_{\text{t}}(\mathbf{r}_j)+\theta_{\text{b}}(\mathbf{r}_j)] \nonumber\\
&-&\gamma\sum_{\langle ij\rangle}\cos[\theta_{\text{t}}(\mathbf{r}_i)-\theta_{\text{t}}(\mathbf{r}_j)]+\cos[\theta_{\text{b}}(\mathbf{r}_i)-\theta_{\text{b}}(\mathbf{r}_j)] \nonumber\\
&+& A\sum_{i}\cos[2\theta_{\text{t}}(\mathbf{r}_i)-2\theta_{\text{b}}(\mathbf{r}_i)].
\end{eqnarray}
Here $\langle ij\rangle$ represents nearest-neighbor bonding, and the positive coefficients $\alpha$, $\lambda$ and $\gamma$ satisfy,
\begin{equation}\label{relation}
\alpha=\frac{\rho-2\gamma}{4},~~~~~~~~ \lambda=\frac{\kappa-2\gamma}{4}.
\end{equation}
Note that although different $\alpha$, $\lambda$ and $\gamma$ satisfying Eq. (\ref{relation}) lead to the same continuous Hamiltonian (\ref{Hamiltonian_r}) in the continuum limit, it is required that all of them should be positive so as to reproduce the correct low-energy ``classical Hilbert space'' for allowed vortices. The reason is as follow. Here the $\alpha>0$ and $\lambda>0$ terms energetically allow for integer or half-integer $\theta_{+}$ and $\theta_{-}$ vortices, while the $\gamma>0$ term energetically only allows for integer $\theta_{\text{t}}$ or $\theta_{\text{b}}$ vortices and hence imposes the ``kinematic constraint'' between the $\theta_{+}$ and $\theta_{-}$ vortices. Note that although the $\gamma$ term does not naturally emerge from Eq. (\ref{Hamiltonian_r}), the singlevaluedness of the $\psi_{\text{t/b}}$ field dictates it. This term is crucial to yield the correct topology of the phase diagram. As shown in the SM~\cite{SM}, if we turn off the $\gamma$ term, $\theta_{+}$ and $\theta_{-}$ are decoupled, leading to topologically wrong phase diagram. A comparison between the correct phase diagram and this wrong one shows that the kinematic correlation makes the vestigial phase regimes largely shrink. For thermodynamic limit, even an infinitesimal $\gamma$ can energetically guarantee the ''kinematic constraint''. Here in the discrete lattice, we set $\gamma=\frac{1}{4}\rho\kappa/(\rho+\kappa),~A=0.025\rho$, and their other values lead to similar results~\cite{SM}.

\begin{figure}[h]
	\centering
	\includegraphics[width=0.45\textwidth]{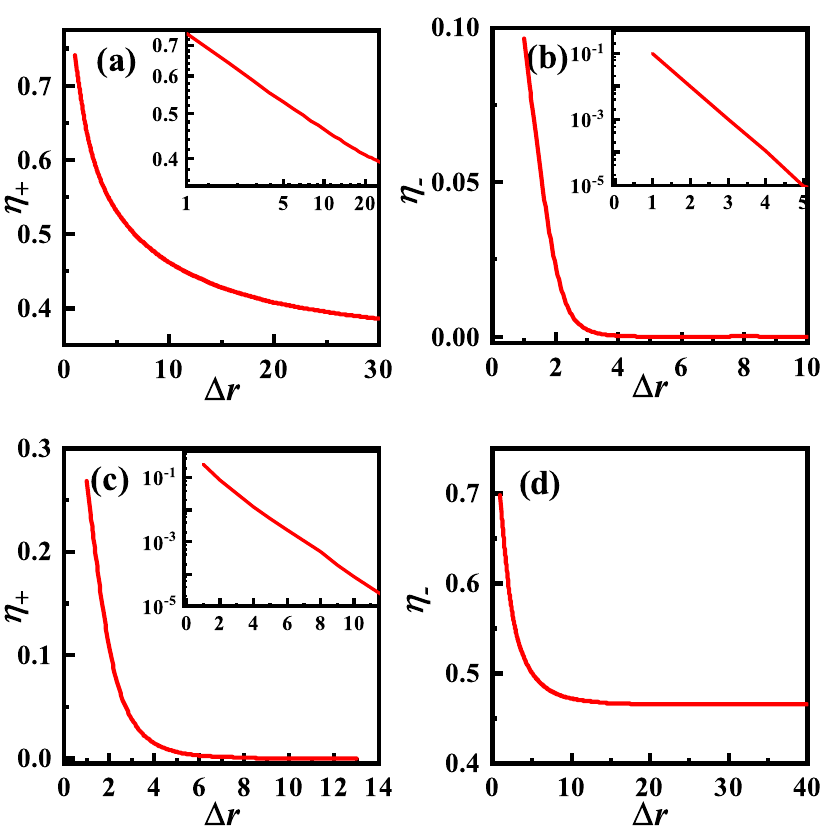}
	\caption{(Color online) The correlation function $\eta_{\pm}$ for (a) and (b) for A($\kappa=0.2, T=0.2$), and for (c) and (d) for D($\kappa=2.2, T=0.5$) marked in Fig.~\ref{phase_diagram}(b). Insets of (a) the log-log plot, and (b) and (c) only the y- axes are logarithmic.}\label{AD}
\end{figure}

The MC phase diagram shown in Fig.~\ref{phase_diagram}(b) is qualitatively consistent with the RG one shown in Fig.~\ref{phase_diagram}(a). Various $T$ dependent quantities are shown in Fig.~\ref{temperature_dependence} for $\kappa/\rho=0.3,~1,~2.2$ marked in Fig.~\ref{phase_diagram}(b), with the formulas adopted in the MC calculations provided in {\bf Methods}. For $\kappa/\rho=0.3$, the specific heat $C_v$ is shown in Fig.~\ref{temperature_dependence}(a), where the high-$T$ broad hump characterizes the Kosterlitz-Thouless (K-T) phase transition between the normal state and the charge-4e SC and the low-$T$ sharp peak characterizes the Ising phase transition between the charge-4e SC and the chiral SC. For this $\kappa/\rho$, Fig.~\ref{temperature_dependence}(b) show the phase stiffness $S$ characterizing the SC and the Ising ODP $I$ characterizing the relative-phase order\cite{SM}, which emerge at the critical temperatures corresponding to the broad hump and sharp peak in Fig.~\ref{temperature_dependence}(a) respectively.  Furthermore, the total- ($\chi_+$) and relative- ($\chi_-$) phase susceptibilities\cite{SM} shown in Fig.~\ref{temperature_dependence}(c) diverge at the same critical temperatures. For $\kappa/\rho=1$, the specific heat shown in Fig.~\ref{temperature_dependence}(d) exhibits only one peak, suggesting a direct phase transition from the normal state to the chiral SC. Such a result is also reflected in Fig.~\ref{temperature_dependence}(e) and (f) which show that the total- and relative- phase (quasi) orders emerge at the same temperature. For $\kappa/\rho=2.2$, the corresponding results shown in Fig.~\ref{temperature_dependence}(g), (h) and (i) reveal that following the decrease of $T$, the system will successively experience the normal state, the chiral metal, and the chiral TSC phases. The results presented in Fig.~\ref{temperature_dependence} are well consistent with the phase diagram shown in Fig.~\ref{phase_diagram}(b).

The total- ($+$) and relative- ($-$) phase correlation functions $\eta_{\pm}$ are shown in Fig.~\ref{AD}. See their formulas in {\bf Methods}. Fig.~\ref{AD}(a) and (b) show that for the representative point A marked in Fig.~\ref{phase_diagram}(b), while $\eta_{+}(\Delta\mathbf{r})$ power-law decays with $\Delta r$ suggesting quasi-long-range order of the total phase, $\eta_{-}(\Delta\mathbf{r})$ decays exponentially with $\Delta r$, suggesting disorder of the relative phase. Obviously, these electron correlations are consistent with the charge-4e SC phase. Fig.~\ref{AD}(c) and (d) show that for the point D, while $\eta_{+}(\Delta\mathbf{r})$ decays exponentially with $\Delta r$ suggesting disorder of the total phase, $\eta_{-}(\Delta\mathbf{r})$ saturates to a constant number for large enough $\Delta r$ suggesting long-range order of the relative phase, consistent with the chiral-metal phase. For comparison, the $\eta_{\pm}$ for the points B and C provided in the SM~\cite{SM} are also consistent with the normal-metal and chiral-SC phases.

~~~~

\noindent{{\bf Discussions}}

~~~~~~

In comparison with previous proposals for the charge-4e/6e SC based on melting of the PDW\cite{Berg2009, Agterberg2011,You2012} or the nematic pairing\cite{Fernandes2021, Jian2021}, our proposal is based on a more definite and easily realized start point: here we only need to start from non-topological $d$-wave SC (or $f$-wave SC) in any four-fold (or six-fold) symmetric monolayers. Particularly, we have provided concrete synthesized materials to realize our proposal, i.e. the 45$^o$-twisted bilayer cuprates and the 30$^o$-twisted bilayer of some graphene family. Further more, superior to previous bilayer approach, here a Cooper pair from the top layer pairs with a Cooper pair from the bottom layer to form the charge-4e SC between the layers. Consequently, the half flux quantization can be experimentally detected as a hallmark of the charge-4e SC in our proposal.

The TB-QC provides a better platform to realize the vestigial phases than conventional chiral superconductors such as the $p+ip$ or $d+id$ ones on the square or honeycomb lattices. The latter also host two degenerate pairing ODPs, and hence can accommodate both total and relative- phase fluctuations of the two ODPs. However, the rotational symmetry of the monolayer system is not as high as that of the TB-QC studied here. Consequently, for chiral TSC in monolayers systems, there can be many nonzero coefficients in Eq. (\ref{Hamiltonian_k}). Particularly, the two phase fields are generally dynamically coupled as the symmetries in these systems allow for extra terms such as $\nabla_{\pm}\theta_{+}\cdot \nabla_{\pm}\theta_{-}$ in the Hamiltonian density in Eq. (\ref{Hamiltonian_r}). See more details in the SM~\cite{SM}. As shown in Fig.~\ref{phase_diagram} and Fig. S5(a), the kinematic correlation between $\theta_{+}$ and $\theta_{-}$ has already made the vestigial phase regimes largely shrink, their extra dynamic coupling might make them further shrink or even vanish.

In conclusion, we have predicted realization of the charge-4e SC or the chiral metal in the TB-QC, emerging as the unilateral (quasi) ordering of the total- or relative- pairing phase of the two layers, above the chiral-TSC ground state. The TB-QC provides a better platform to realize these vestigial phases than previous proposals as here we can start from a more definite and easily realized start point.

~~~~

\noindent{{\bf Methods}}

~~~~~~

\noindent{{\bf The RG Approach:}} Here we provide some technique details for the RG study. With standard RG analysis, the flow equations at the one-loop level are given by:
\begin{eqnarray}
\label{eqn:RG-equation}
\frac{dg_{2,0}}{d\ln b}&=&(2-\pi\rho^{'})g_{2,0} \nonumber \\
\frac{dg_{0,2}}{d\ln b}&=&(2-\pi\kappa^{'})g_{0,2} \nonumber \\
\frac{dg_{1,1}}{d\ln b}&=&\left(2-\frac{\pi}{4}(\rho^{'}+\kappa^{'})\right)g_{1,1} \nonumber \\
\frac{dg_{4}}{d\ln b}&=&(2-\frac{4}{\pi\kappa^{'}})g_{4} \nonumber \\
\frac{d\rho^{'}}{d\ln b}&=&-16g_{2,0}^{2}\rho^{'3}-\frac{g_{1,1}^{2}}{2}\rho^{'2}(\rho^{'}+\kappa^{'}) \nonumber \\
\frac{d\kappa^{'}}{d\ln b}&=&\frac{256g_{4}^{2}}{\pi^4\kappa^{'}}-16g_{0,2}^{2}\kappa^{'3}-\frac{g_{1,1}^{2}}{2}\kappa^{'2}(\rho^{'}+\kappa^{'}),
\end{eqnarray}
Here $b$ represents the renormalization scale, $g_{2,0}$, $g_{0,2}$ and $g_{1,1}$ represent the coupling strength of different types of topological defects, $\rho^{'}=\rho/T$ and $\kappa^{'}=\kappa/T$ represent two kinds of stiffness parameters.

\begin{table}[!h]
\centering
\caption{Fixed points of the coupling parameters under RG, and the corresponding phases.}\label{tab:1}
\begin{tabular}{|c|c|c|c|c|}
  \hline\hline
  $g_{2,0}$ & $g_{0,2}$ & $g_{4}$ & $g_{1,1}$ & phase \\
  \hline
  $\infty$ & $\infty$ & 0 & $\infty$ & normal \\
  \hline
  $\infty$ & 0 & 0 & $\infty$ & normal \\
  \hline
  0 & 0 & 0 & $\infty$ & normal \\
  \hline
  0 & $\infty$ & 0 & $\infty$ & normal \\
  \hline
  $\infty$ & $\infty$ & 0 & 0 & normal \\
  \hline
  0 & $\infty$ & 0 & 0 & charge 4e SC \\
  \hline
  0 & 0 & $\infty$ & 0 & chiral SC \\
  \hline
  $\infty$ & 0 & $\infty$ & 0 & chiral metal \\
  \hline\hline
\end{tabular}
\end{table}

In Table~\ref{tab:1}, we present eight possible fixed points of the RG flow equation (\ref{eqn:RG-equation}) and the corresponding phases. We have not listed the renormalized values of the stiffness parameters ($\rho^{'}$ and $\kappa^{'}$), because they are consistent with the phase revealed by the RG flow result of the g-couplings. Specifically, the $\rho^{'}$ flows to a finite positive value if the U(1)-gauge symmetry is (quasi) broken, otherwise it flows to zero; the $\kappa^{'}$ flows to infinity if the time- reverse symmetry is broken, otherwise it flows to zero. See more details in the SM~\cite{SM}. In addition, although five possible flow results for the normal state are listed in the table, only the first one actually appears in our calculations. Furthermore, following the standard process \cite{Park2021}, we also provide stability analysis of the fixed points in the SM \cite{SM}.

~~~~~~~~~~~~~~~~~~~~~~

\noindent{{\bf The Monte-Carlo Approach:}} Here we provide some formula for the MC calculations.

The phase stiffness characterizing the quasi-long-range order of the total-phase and hence the SC is \cite{Zeng2021}
\begin{eqnarray}\label{ss}
	S=\frac{1}{N}(<H_x>-\beta<I_x^2>)
\end{eqnarray}
with
\begin{eqnarray}\label{Hamiltonian_p1}
H_x &=& 4\alpha\sum_{<ij>_x}\cos[\theta_{\text{t}}(\mathbf{r}_i)+\theta_{\text{b}}(\mathbf{r}_i)-\theta_{\text{t}}(\mathbf{r}_j)+\theta_{\text{b}}(\mathbf{r}_j)]\nonumber\\
&&+\gamma\sum_{<ij>_x}\cos[\theta_{\text{t}}(\mathbf{r}_{i})-\theta_{\text{t}}(\mathbf{r}_{j})]+\cos[\theta_{\text{b}}(\mathbf{r}_{i})-\theta_{\text{b}}(\mathbf{r}_{j})]\nonumber\\
I_x &=& 2\alpha\sum_{<ij>_x}\sin[\theta_{\text{t}}(\mathbf{r}_i)+\theta_{\text{b}}(\mathbf{r}_i)-\theta_{\text{t}}(\mathbf{r}_j)+\theta_{\text{b}}(\mathbf{r}_j)]\nonumber\\&&+\gamma\sum_{<ij>_x}\sin[\theta_{\text{t}}(\mathbf{r}_{i})-\theta_{\text{t}}(\mathbf{r}_{j})]+\sin[\theta_{\text{b}}(\mathbf{r}_{i})-\theta_{\text{b}}(\mathbf{r}_{j})],\nonumber\\
\end{eqnarray}
where $N$ is the site number, and $\beta=1/k_BT$.

The Ising order parameter characterizing the relative-phase ordering breaking the time-reversal symmetry is,
\begin{equation}
I\equiv\frac{1}{N^2}\sum_{ij}\left\langle\sin[\theta_{\text{t}}(\mathbf{r}_{i})-\theta_{\text{b}}(\mathbf{r}_{i})]\cdot \sin[\theta_{\text{t}}(\mathbf{r}_{j})-\theta_{\text{b}}(\mathbf{r}_{j})]\right\rangle.\nonumber
\end{equation}

The total- ($+$) and relative- ($-$) phase susceptibilities for temperatures above the $T_c$ of the corresponding orders are defined by
\begin{equation}
\chi_{\pm}\equiv \frac{1}{NT}\sum_{i}\left\langle\left| e^{i\left[\theta_{\text{t}}\left(\mathbf{r}_i\right)\pm\theta_{\text{b}}\left(\mathbf{r}_i\right)\right]}\right|^2\right\rangle.
\end{equation}

The total- ($+$) and relative- ($-$) phase correlation functions are defined as
\begin{equation}
\eta_{\pm}(\Delta \mathbf{r})=\frac{1}{N}\sum_{\mathbf{r}}\left\langle e^{i[\theta_{t}(\mathbf{r})\pm\theta_{b}(\mathbf{r})-\theta_{t}(\mathbf{r}+\Delta \mathbf{r})\mp\theta_{b} (\mathbf{r}+\Delta \mathbf{r})]}\right\rangle.
\end{equation}

~~~~~~~~~~~~~~~~~~~~~~~~~~~~~~~~~~~

{\it Acknowledgements:} We are grateful to the stimulating discussions with Zhi-Ming Pan, Shao-Kai Jian, Chen Lu, Meng Zeng and Wei-Qiang Chen. This work is supported by the NSFC under the Grant Nos. 12074031, 12234016, 12174317, 11674025.

~~~~~~~~~~~~~~~~~~~~~~~~~~~~~~~~~~~~~~~

{\it Data Availability:}
All data are displayed in the main text and Supplementary Information.

~~~~~~~~~~~~~~~~~~~~~~~~~~~~~~~~~~~~~

{\it Code Availability:}
The code that supports the plots within this paper are available from the corresponding author upon reasonable request.

\newpage

\renewcommand{\theequation}{S\arabic{equation}}
\setcounter{equation}{0}
\renewcommand{\thefigure}{S\arabic{figure}}
\setcounter{figure}{0}
\renewcommand{\thetable}{S\arabic{table}}
\setcounter{table}{0}
\begin{widetext}

\appendix

\section{Derivation of the effective Hamiltonian from Ginzburg-Landau theory}
In this section, we derive the effective Hamiltonian appearing in the Eq. (12) of the main text by expanding the Ginzburg-Landau (G-L) free energy up to the fourth-order term of the order parameters.

\subsection{Symmetry}
To elucidate the effect of the symmetry operations on the argument of the G-L free-energy functional, let's start from the mean-field BCS Hamiltonian:
\begin{eqnarray}
\label{eqn:BCS-Hamiltonian}
H_{\text{BCS-MF}}=H_{\text{TB}}+\sum_{\mathbf{r},\delta}c_{\mathbf{r},\text{t}\uparrow}^{\dagger}c_{\mathbf{r}+\delta,\text{t}\downarrow}^{\dagger}\Gamma^{(\text{t})}(\delta)\psi_{\text{t}}(\mathbf{r})+
c_{\mathbf{r},\text{b}\uparrow}^{\dagger}c_{\mathbf{r}+\delta,\text{b}\downarrow}^{\dagger}\Gamma^{(\text{b})}(\delta)\psi_{\text{b}}(\mathbf{r})+h.c.
\end{eqnarray}
Here $\mathbf{r}$ labels the center-of-mass coordinate of a Cooper pair and $\delta$ is the relative coordinate between the two electrons within a Cooper pair. $\Gamma^{(\mu)}(\delta)$ is the fixed normalized real form factor with $\mu=\text{t/b}$, and $\psi_{\mu}(\mathbf{r})$ is a slowly-varying ``envelope'' function describing the spatial fluctuation of the complex pairing amplitude at finite temperature.  Each symmetry operation first acts on the $c$ and $c^{\dagger}$ operators, then through a dummy-index transformation, the effect is transferred to the action of the $\Gamma$ and $\psi$. As the $\Gamma$ has simple transformation rule under the symmetry, i.e. it changes sign upon every $C_{n}^1$ operation and changes or does not change upon the mirror reflection operation, the effect can be transferred purely to $\psi$. Therefore, we have chosen an gauge in which each symmetry operation only acts on $\psi_{\mu}(\mathbf{r})$.

Under $\widetilde{C}_{2n}^{1}$, the spatial dependent pairing amplitudes change to:
\begin{eqnarray}
\psi_{\text{b}}(\mathbf{r})\to\widetilde{\psi}_{\text{b}}(\mathbf{r})=\psi_{\text{t}}(\widehat{P}_{\frac{\pi}{n}}^{-1}\mathbf{r}),\qquad \psi_{\text{t}}(\mathbf{r})\to\widetilde{\psi}_{\text{t}}(\mathbf{r})=-\psi_{\text{b}}(\widehat{P}_{\frac{\pi}{n}}^{-1}\mathbf{r}).
\end{eqnarray}
Under the mirror reflection operation $\widehat{P}$, it is easy to prove (we have chosen a gauge without loss of generality):
\begin{eqnarray}
\psi_{\text{b}}(\mathbf{r})\to\widetilde{\psi}_{\text{b}}(\mathbf{r})=-\psi_{\text{b}}(\widehat{P}^{-1}\mathbf{r}),\qquad
\psi_{\text{t}}(\mathbf{r})\to\widetilde{\psi}_{\text{t}}(\mathbf{r})=\psi_{\text{t}}(\widehat{P}^{-1}\mathbf{r}).
\end{eqnarray}
For convenience, we rotate the basis to $\psi_{\pm}=\psi_{t}\pm i\psi_{b}$ and rewrite the above transformation in the $\mathbf{k}-$space
\begin{eqnarray}
\psi_{+}(\mathbf{k}) \xrightarrow{\widetilde{C}_{2n}^{1}} e^{i\pi/2}\psi_{+}(\widehat{P}_{\frac{\pi}{n}}^{-1}\mathbf{k}),&~&\qquad \psi_{-}(\mathbf{k}) \xrightarrow{\widetilde{C}_{2n}^{1}}e^{-i\pi/2}\psi_{-}(\widehat{P}_{\frac{\pi}{n}}^{-1}\mathbf{k})\nonumber\\
\psi_{+}(\mathbf{k})\xrightarrow{\widehat{P}}\psi_{-}(\widehat{P}^{-1}\mathbf{k}),&~&\qquad
\psi_{-}(\mathbf{k})\xrightarrow{\widehat{P}}\psi_{+}(\widehat{P}^{-1}\mathbf{k}).
\end{eqnarray}
Here we consider the $\widetilde{C}_{2n}^{1}$ and the mirror reflection, but neglect the time-reversal symmetry. The final effect of the time-reversal symmetry on the Hamiltonian is consistent with that obtained with only considering the $\widetilde{C}_{2n}^{1}$ and the mirror reflection symmetries.

With the definition $\mathbf{k}_{\pm}=k_{x}\pm ik_{y}$, we obtain the momentum transformation relations:
\begin{eqnarray}
\widehat{P}_{\frac{\pi}{n}}\mathbf{k}_{+}=e^{i\pi/n}\mathbf{k}_{+},\qquad \widehat{P}_{\frac{\pi}{n}}\mathbf{k}_{-}=e^{-i\pi/n}\mathbf{k}_{-}.
\end{eqnarray}

\subsection{The second-order G-L expansion}
Up to the lowest-order expansion, the differential term in G-L free energy has the following general form in the $\mathbf{k}-$ space:
\begin{eqnarray}
F_{0}^{(2)}& = &\sum_{\mathbf{k}}\psi_{+}^{\ast}(\mathbf{k})\psi_{+}(\mathbf{k})(a_{1}\mathbf{k}_{+}^2+b_{1}\mathbf{k}_{-}^2+c_{1}\mathbf{k}_{+}\mathbf{k}_{-}) \nonumber\\
& + &\sum_{\mathbf{k}}\psi_{+}^{\ast}(\mathbf{k})\psi_{-}(\mathbf{k})(a_{2}\mathbf{k}_{+}^2+b_{2}\mathbf{k}_{-}^2+c_{2}\mathbf{k}_{+}\mathbf{k}_{-}) \nonumber\\
& + &\sum_{\mathbf{k}}\psi_{-}^{\ast}(\mathbf{k})\psi_{+}(\mathbf{k})(a_{3}\mathbf{k}_{+}^2+b_{3}\mathbf{k}_{-}^2+c_{3}\mathbf{k}_{+}\mathbf{k}_{-}) \nonumber\\
& + &\sum_{\mathbf{k}}\psi_{-}^{\ast}(\mathbf{k})\psi_{-}(\mathbf{k})(a_{4}\mathbf{k}_{+}^2+b_{4}\mathbf{k}_{-}^2+c_{4}\mathbf{k}_{+}\mathbf{k}_{-}).
\end{eqnarray}
Under the operation $\widetilde{C}_{2n}^{1}$, $F_{0}^{(2)}$ change to:
\begin{eqnarray}
\label{eqn:F2_transform}
F_{0}^{(2)}\xrightarrow{\widetilde{C}_{2n}^{1}} &=&\sum_{\mathbf{k}}\psi_{+}^{\ast}(\mathbf{k})\psi_{+}(\mathbf{k})(a_{1}e^{i2\pi/n}\mathbf{k}_{+}^2+b_{1}e^{-i2\pi/n}\mathbf{k}_{-}^2+c_{1}\mathbf{k}_{+}\mathbf{k}_{-})\nonumber\\
&+&\sum_{\mathbf{k}}e^{-i2\pi/2}\psi_{+}^{\ast}(\mathbf{k})\psi_{-}(\mathbf{k})(a_{2}e^{i2\pi/n}\mathbf{k}_{+}^2+b_{2}e^{-i2\pi/n}\mathbf{k}_{-}^2+c_{2}\mathbf{k}_{+}\mathbf{k}_{-})\nonumber\\
&+&\sum_{\mathbf{k}}e^{i2\pi/2}\psi_{-}^{\ast}(\mathbf{k})\psi_{+}(\mathbf{k})(a_{3}e^{i2\pi/n}\mathbf{k}_{+}^2+b_{3}e^{-i2\pi/n}\mathbf{k}_{-}^2+c_{3}\mathbf{k}_{+}\mathbf{k}_{-})\nonumber\\
&+&\sum_{\mathbf{k}}\psi_{-}^{\ast}(\mathbf{k})\psi_{-}(\mathbf{k})(a_{4}e^{i2\pi/n}\mathbf{k}_{+}^2+b_{4}e^{-i2\pi/n}\mathbf{k}_{-}^2+c_{4}\mathbf{k}_{+}\mathbf{k}_{-}).
\end{eqnarray}
As $n=4$ or $6$, the invariance of $F_{0}^{(2)}$ requires only $c_{1},c_{4}\neq0$ while all the other coefficients keep zero. Further more, $c_{1}=c_{4}=B$ is required by the mirror-reflection symmetry. Changing back to the real space, we arrive at the form of $F_{0}^{(2)}$ as following:
\begin{eqnarray}
\label{eqn:F2}
F_{0}^{(2)}&=&B\int d^{2}\mathbf{r}[(\nabla\psi_{+}^{\ast})\cdot(\nabla\psi_{+})+(\nabla\psi_{-}^{\ast})\cdot(\nabla\psi_{-})] \nonumber\\
&=&B\psi_{0}^{2}\int d^{2}\mathbf{r}[\nabla(e^{-i\theta_{t}}-ie^{-i\theta_{b}})\cdot\nabla(e^{i\theta_{t}}+ie^{i\theta_{b}})+\nabla(e^{-i\theta_{t}}
+ie^{-i\theta_{b}})\cdot\nabla(e^{i\theta_{t}}-ie^{i\theta_{b}})] \nonumber\\
&=&2B\psi_{0}^{2}\int d^{2}\mathbf{r}[(\nabla\theta_{b})^{2}+(\nabla\theta_{t})^{2}] \nonumber\\
&=&4B\psi_{0}^{2}\int d^{2}\mathbf{r}[|\nabla\theta_{+}|^{2}+|\nabla\theta_{-}|^{2}]
\end{eqnarray}
Here $\psi_0$ represents the amplitude of the pairing order parameter.

\subsection{The fourth-order G-L expansion}
According to the second order expansion of the differential term in the G-L free energy, the coefficients before $\theta_{+}$ and $\theta_{-}$ are the same. To get different coefficients, we need expand $F_{0}$ to the fourth order with the general form as:
\begin{eqnarray}
F_{0}^{(4)}&=&\sum_{\mathbf{k}_{1},\mathbf{k}_{2},\mathbf{k}_{3},\mathbf{k}_{4}}\psi_{\alpha}^{\ast}(\mathbf{k}_{1})\psi_{\beta}^{\ast}(\mathbf{k}_{2})\psi_{\gamma}(\mathbf{k}_{3})\psi_{\nu}(\mathbf{k}_{4})
(\sum_{i,j=1}^{4}\alpha_{ij}\mathbf{k}_{+i}\cdot\mathbf{k}_{+j}+\beta_{ij}\mathbf{k}_{-i}\cdot\mathbf{k}_{-j}+\gamma_{ij}\mathbf{k}_{+i}\cdot\mathbf{k}_{-j}+\nu_{ij}\mathbf{k}_{-i}\cdot\mathbf{k}_{+j}) \nonumber\\
\end{eqnarray}
where $\alpha,\beta,\gamma,\nu=\pm$. It is easy to verify that $\alpha+\beta+\gamma+\nu$ should be an even integer. Since the angular momentum of $\psi_{\alpha}^{(\ast)}(\mathbf{k}_{i})$ is $\pm n/2$, that of $\psi_{\alpha}^{\ast}(\mathbf{k}_{1})\psi_{\beta}^{\ast}(\mathbf{k}_{2})\psi_{\gamma}(\mathbf{k}_{3})\psi_{\nu}(\mathbf{k}_{4})$ should be an integer times $n$. And the angular momentum of $k_{\pm}$ is $\pm1$. The invariance of $F_{0}^{(4)}$ under $\widetilde{C}_{2n}^{1}$ requires that the total angular momentum should be zero. For $n=4$ or $6$, the restriction of zero total angular momentum dictates $\alpha_{ij}=\beta_{ij}=0$. Then, we can simplify the general form of $F_{0}^{(4)}$:
\begin{eqnarray}
\label{eqn:genergal_F4}
F_{0}^{(4)}&=&\sum_{\mathbf{k}_{1},\mathbf{k}_{2},\mathbf{k}_{3},\mathbf{k}_{4}}\psi_{+}^{\ast}(\mathbf{k}_{1})\psi_{+}^{\ast}(\mathbf{k}_{2})\psi_{+}(\mathbf{k}_{3})\psi_{+}(\mathbf{k}_{4})
(\sum_{i,j=1}^{4}2\gamma^{(1)}_{ij}\mathbf{k}_{i}\cdot \mathbf{k}_{j}) \nonumber\\
&+&\sum_{\mathbf{k}_{1},\mathbf{k}_{2},\mathbf{k}_{3},\mathbf{k}_{4}}\psi_{+}^{\ast}(\mathbf{k}_{1})\psi_{-}^{\ast}(\mathbf{k}_{2})\psi_{+}(\mathbf{k}_{3})\psi_{-}(\mathbf{k}_{4})
(\sum_{i,j=1}^{4}2\gamma^{(2)}_{ij}\mathbf{k}_{i}\cdot \mathbf{k}_{j}) \nonumber\\
&+&\sum_{\mathbf{k}_{1},\mathbf{k}_{2},\mathbf{k}_{3},\mathbf{k}_{4}}\psi_{-}^{\ast}(\mathbf{k}_{1})\psi_{-}^{\ast}(\mathbf{k}_{2})\psi_{-}(\mathbf{k}_{3})\psi_{-}(\mathbf{k}_{4})
(\sum_{i,j=1}^{4}2\gamma^{(3)}_{ij}\mathbf{k}_{i}\cdot \mathbf{k}_{j})
\end{eqnarray}

We now consider the first and the third term in the general form of $F_{0}^{(4)}$ since there is only $\psi_{+}$ or $\psi_{-}$. It is easy to verify that $\psi_{\pm}\rightarrow\psi_{\mp}^{*}$ under TRS. Remembering all the transformation relation in mind, the form of equation (\ref{eqn:genergal_F4}) can be further simplified as:
\begin{eqnarray}
\label{eqn:F4_13}
F^{(4)}_{0(1,3)}=\sum_{\mathbf{k}_{1},\mathbf{k}_{2},\mathbf{k}_{3},\mathbf{k}_{4}}[\psi_{+}^{\ast}(\mathbf{k}_{1})\psi_{+}^{\ast}(\mathbf{k}_{2})\psi_{+}(\mathbf{k}_{3})\psi_{+}(\mathbf{k}_{4})  +\psi_{-}^{\ast}(\mathbf{k}_{1})\psi_{-}^{\ast}(\mathbf{k}_{2})\psi_{-}(\mathbf{k}_{3})\psi_{-}(\mathbf{k}_{4})] \nonumber\\
\cdot[a(\mathbf{k}_{1}^{2}+\mathbf{k}_{2}^{2}+\mathbf{k}_{3}^{2}+\mathbf{k}_{4}^{2})+b(\mathbf{k}_{1}\cdot \mathbf{k}_{2}+\mathbf{k}_{3}\cdot \mathbf{k}_{4})+c(\mathbf{k}_{1}+\mathbf{k}_{2})\cdot(\mathbf{k}_{3}+\mathbf{k}_{4})]
\end{eqnarray}
A valuable equation $(\mathbf{k}_{1}+\mathbf{k}_{2}-\mathbf{k}_{3}-\mathbf{k}_{4})^{2}=0$ should be emphasized before the proceeding process. Expanding this equation, we have:
\begin{equation}
\sum_{i=1}^{4}\mathbf{k}_{i}^{2}=2(\mathbf{k}_{1}+\mathbf{k}_{2})\cdot(\mathbf{k}_{3}+\mathbf{k}_{4})-2(\mathbf{k}_{1}\cdot \mathbf{k}_{2}+\mathbf{k}_{3}\cdot \mathbf{k}_{4}).
\end{equation}
We can rewrite the first and third term:
\begin{eqnarray}
F^{(4)}_{0(1,3)}=\sum_{\mathbf{k}_{1},\mathbf{k}_{2},\mathbf{k}_{3},\mathbf{k}_{4}}[\psi_{+}^{\ast}(\mathbf{k}_{1})\psi_{+}^{\ast}(\mathbf{k}_{2})\psi_{+}(\mathbf{k}_{3})\psi_{+}(\mathbf{k}_{4})  +\psi_{-}^{\ast}(\mathbf{k}_{1})\psi_{-}^{\ast}(\mathbf{k}_{2})\psi_{-}(\mathbf{k}_{3})\psi_{-}(\mathbf{k}_{4})] \nonumber\\
\cdot[ (b-2a)\cdot(\mathbf{k}_{1}\cdot \mathbf{k}_{2}+\mathbf{k}_{3}\cdot \mathbf{k}_{4}) + (c+2a)\cdot(\mathbf{k}_{1}+\mathbf{k}_{2})\cdot(\mathbf{k}_{3}+\mathbf{k}_{4}) ]
\end{eqnarray}

By the same method, the second term of the fourth order expansion of the differential term in G-L free energy is
\begin{eqnarray}
\label{eqn:F4_2}
F^{(4)}_{0(2)}&=&\sum_{\mathbf{k}_{1},\mathbf{k}_{2},\mathbf{k}_{3},\mathbf{k}_{4}}\psi_{+}^{\ast}(\mathbf{k}_{1})\psi_{-}^{\ast}(\mathbf{k}_{2})\psi_{+}(\mathbf{k}_{3})\psi_{-}(\mathbf{k}_{4}) \nonumber\\
&\cdot&[a'(\mathbf{k}_{1}^{2}+\mathbf{k}_{2}^{2}+\mathbf{k}_{3}^{2}+\mathbf{k}_{4}^{2})+b'(\mathbf{k}_{1}\cdot\mathbf{k}_{2}+\mathbf{k}_{3}\cdot \mathbf{k}_{4})
+c'(\mathbf{k}_{1}\cdot\mathbf{k}_{3}+\mathbf{k}_{2}\cdot \mathbf{k}_{4})+d'(\mathbf{k}_{1}\cdot\mathbf{k}_{4}+\mathbf{k}_{2}\cdot \mathbf{k}_{3}) ]
\end{eqnarray}
Transforming to the real space, the total form of $F_{0}^{(4)}$ is:
\begin{eqnarray}
F^{(4)}&=&-(b-2a)\int d^{2}\mathbf{r}[ (\nabla\psi_{+}^{\ast})^{2}\psi_{+}^{2}+(\psi_{+}^{\ast})^{2}(\nabla\psi_{+})^{2} + (\nabla\psi_{-}^{\ast})^{2}\psi_{-}^{2}+(\psi_{-}^{\ast})^{2}(\nabla\psi_{-})^{2}  ] \nonumber\\
&+&(c+2a)\int d^{2}\mathbf{r}[ \nabla(\psi_{+}^{\ast 2})\cdot\nabla(\psi_{+}^{2})+\nabla(\psi_{-}^{\ast 2})\cdot\nabla(\psi_{-}^{2}) ] \nonumber\\
&-&(b'-2a')\int d^{2}\mathbf{r}[ (\nabla\psi_{+}^{\ast})\cdot(\nabla\psi_{-}^{\ast})\psi_{+}\psi_{-}+\psi_{+}^{\ast}\psi_{-}^{\ast}(\nabla\psi_{+})\cdot(\nabla\psi_{-})]\nonumber\\
&+&(c'+2a')\int d^{2}\mathbf{r}[ \nabla\psi_{+}^{\ast}\cdot\nabla\psi_{+}|\psi_{-}|^{2}+|\psi_{+}|^{2}\nabla\psi_{-}^{\ast}\cdot\nabla\psi_{-} ]\nonumber\\
&+&(d'+2a')\int d^{2}\mathbf{r} [ \nabla\psi_{+}^{\ast}\cdot\nabla\psi_{-}\psi_{-}^{\ast}\psi_{+} + \nabla\psi_{-}^{\ast}\cdot\nabla\psi_{+}\psi_{+}^{\ast}\psi_{-} ] \nonumber\\
&=&32(b+2c+2a)\psi_{0}^{4} \int d^{2}\mathbf{r}|\nabla\theta_{+}|^{2}+16(c'+2a')\psi_{0}^{4}\int d^{2}\mathbf{r}|\nabla\theta_{-}|^{2}.
\end{eqnarray}

So, the stiffness parameters $\rho$ and $\kappa$ in the text take the form as:
\begin{eqnarray}
\rho&=&8B\psi_{0}^{2}+64(b+c)\psi_{0}^{4}, \\
\kappa&=&8B\psi_{0}^{2}+32(c'+2a')\psi_{0}^{4}.
\end{eqnarray}
And the the lowest order of the real space Hamiltonian is given by:
\begin{equation}
H_{0}=\int d^{2}\mathbf{r}\left( \frac{\rho}{2}|\nabla\theta_{+}|^{2}+\frac{\kappa}{2}|\nabla\theta_{-}|^{2} \right)
\end{equation}

	\section{Derivation of the effective Hamiltonian of chiral TSC in monolayers systems from G-L theory}
	In this section, as the contrast to the previous section, we derive the effective Hamiltonian of chiral TSC in monolayers systems by expanding the G-L free energy up to the second-order term of the order parameters. In the following derivation, we take the $d+id$ TSC in the hexagonal lattice as an example. The final result shows that there are extra dynamic couplings of $\theta_+$ and $\theta_-$.
	
	\subsection{Symmetry}
	To elucidate the effect of the symmetry operations on the argument of the G-L free-energy functional, let's start from the mean-field BCS Hamiltonian:
	\begin{eqnarray}
		\label{eqn:BCS-Hamiltonian1}
		H_{\text{BCS-MF}}=H_{\text{TB}}+\sum_{\mathbf{r},\delta}c_{\mathbf{r}\uparrow}^{\dagger}c_{\mathbf{r}+\delta\downarrow}^{\dagger}\Gamma^{(1)}(\delta)\psi_{1}(\mathbf{r})+
		c_{\mathbf{r}\uparrow}^{\dagger}c_{\mathbf{r}+\delta\downarrow}^{\dagger}\Gamma^{(2)}(\delta)\psi_{2}(\mathbf{r})+h.c.
	\end{eqnarray}
	Here 1 and 2 mark the two degenerate components of the d-wave SC. The interpretation of the Eq. (\ref{eqn:BCS-Hamiltonian1}) is the same as the Eq. (\ref{eqn:BCS-Hamiltonian}) except that the component index $1,2$ replaces the layer index $t,b$.
	
	For convenience, we rotate the basis to $\psi_{\pm}=\psi_{1}\pm i\psi_{2}$ and rewrite the above transformation in the $\mathbf{k}-$space. Under $C_{6}^{1}$, the spatial dependent pairing amplitudes change to:
	\begin{eqnarray}
		\psi_{+}(\mathbf{k})\to\widetilde{\psi}_{+}(\mathbf{k})=e^{i2\pi/3}\psi_{+}(\widehat{P}_{\frac{\pi}{3}}^{-1}\mathbf{k}),\qquad \psi_{-}(\mathbf{k})\to\widetilde{\psi}_{-}(\mathbf{k})==e^{-i2\pi/3}\psi_{-}(\widehat{P}_{\frac{\pi}{3}}^{-1}\mathbf{k}).
	\end{eqnarray}
	Under the mirror reflection operation $\widehat{P}$, it is easy to prove (we have chosen a gauge without loss of generality):
	\begin{eqnarray}
		\psi_{+}(\mathbf{k})\to\widetilde{\psi}_{+}(\mathbf{k})=\psi_{-}(\widehat{P}^{-1}\mathbf{k}),\qquad
		\psi_{-}(\mathbf{k})\to\widetilde{\psi}_{-}(\mathbf{k})=\psi_{+}(\widehat{P}^{-1}\mathbf{k}).
	\end{eqnarray}
	
	Here we consider the $C_{6}^{1}$ and the mirror reflection, but neglect the time-reversal symmetry. The final effect of the time-reversal symmetry on the Hamiltonian is consistent with that obtained with only considering the $C_{6}^{1}$ and the mirror reflection symmetries.

	With the definition $\mathbf{k}_{\pm}=k_{x}\pm ik_{y}$, we obtain the momentum transformation relations:
	\begin{eqnarray}
		\widehat{P}_{\frac{\pi}{3}}\mathbf{k}_{+}=e^{i\pi/3}\mathbf{k}_{+},\qquad \widehat{P}_{\frac{\pi}{3}}\mathbf{k}_{-}=e^{-i\pi/3}\mathbf{k}_{-}.
	\end{eqnarray}
	
	\subsection{The second-order G-L expansion}
	Up to the lowest-order expansion, the differential term in G-L free energy has the following general form in the $\mathbf{k}-$ space:
	\begin{eqnarray}
		F_{0}^{(2)}& = &\sum_{\mathbf{k}}\psi_{+}^{\ast}(\mathbf{k})\psi_{+}(\mathbf{k})(a_{1}\mathbf{k}_{+}^2+b_{1}\mathbf{k}_{-}^2+c_{1}\mathbf{k}_{+}\mathbf{k}_{-}) \nonumber\\
		& + &\sum_{\mathbf{k}}\psi_{+}^{\ast}(\mathbf{k})\psi_{-}(\mathbf{k})(a_{2}\mathbf{k}_{+}^2+b_{2}\mathbf{k}_{-}^2+c_{2}\mathbf{k}_{+}\mathbf{k}_{-}) \nonumber\\
		& + &\sum_{\mathbf{k}}\psi_{-}^{\ast}(\mathbf{k})\psi_{+}(\mathbf{k})(a_{3}\mathbf{k}_{+}^2+b_{3}\mathbf{k}_{-}^2+c_{3}\mathbf{k}_{+}\mathbf{k}_{-}) \nonumber\\
		& + &\sum_{\mathbf{k}}\psi_{-}^{\ast}(\mathbf{k})\psi_{-}(\mathbf{k})(a_{4}\mathbf{k}_{+}^2+b_{4}\mathbf{k}_{-}^2+c_{4}\mathbf{k}_{+}\mathbf{k}_{-}).
	\end{eqnarray}
	Under the operation $C_{6}^{1}$, $F_{0}^{(2)}$ change to:
	\begin{eqnarray}
		\label{eqn:F2_transform}
		F_{0}^{(2)}\xrightarrow{C_{6}^{1}} &=&\sum_{\mathbf{k}}\psi_{+}^{\ast}(\mathbf{k})\psi_{+}(\mathbf{k})(a_{1}e^{i2\pi/3}\mathbf{k}_{+}^2+b_{1}e^{-i2\pi/3}\mathbf{k}_{-}^2+c_{1}\mathbf{k}_{+}\mathbf{k}_{-})\nonumber\\
		&+&\sum_{\mathbf{k}}e^{-i4\pi/3}\psi_{+}^{\ast}(\mathbf{k})\psi_{-}(\mathbf{k})(a_{2}e^{i2\pi/3}\mathbf{k}_{+}^2+b_{2}e^{-i2\pi/3}\mathbf{k}_{-}^2+c_{2}\mathbf{k}_{+}\mathbf{k}_{-})\nonumber\\
		&+&\sum_{\mathbf{k}}e^{i4\pi/3}\psi_{-}^{\ast}(\mathbf{k})\psi_{+}(\mathbf{k})(a_{3}e^{i2\pi/3}\mathbf{k}_{+}^2+b_{3}e^{-i2\pi/3}\mathbf{k}_{-}^2+c_{3}\mathbf{k}_{+}\mathbf{k}_{-})\nonumber\\
		&+&\sum_{\mathbf{k}}\psi_{-}^{\ast}(\mathbf{k})\psi_{-}(\mathbf{k})(a_{4}e^{i2\pi/3}\mathbf{k}_{+}^2+b_{4}e^{-i2\pi/3}\mathbf{k}_{-}^2+c_{4}\mathbf{k}_{+}\mathbf{k}_{-}).
	\end{eqnarray}
	the invariance of $F_{0}^{(2)}$ requires only $c_{1},b_{2},a_{3},c_{4}\neq0$ while all the other coefficients keep zero. Further more, $c_{1}=c_{4}=B$ and $b_{2}=a_{3}=C$ are required by the mirror-reflection symmetry. Changing back to the real space, we arrive at the form of $F_{0}^{(2)}$ as following:
	\begin{eqnarray}
		\label{eqn:F2}
		F_{0}^{(2)}&=&\int d^{2}\mathbf{r}B[(\nabla\psi_{+}^{\ast})\cdot(\nabla\psi_{+})+(\nabla\psi_{-}^{\ast})\cdot(\nabla\psi_{-})] \nonumber\\
		&+&
		C[(\nabla_+\psi_{-}^{\ast})\cdot(\nabla_+\psi_{+})+(\nabla_-\psi_{+}^{\ast})\cdot(\nabla_-\psi_{-})] 
	\end{eqnarray}
	Where the C term is the extra dynamic coupling of $\theta_+$ and $\theta_-$. Similar to the above derivation, such extra dynamic coupling is present in all possible chiral TSC ($p+ip$,$d+id$ in hexagonal lattice and $p+ip$ in square lattice) in monolayers systems.

\section{Stability analysis of the fixed points}
By the standard RG analysis, the flow equations at the one-loop level are given by:
\begin{eqnarray}
\label{eqn:RG-equation}
\frac{dg_{2,0}}{d\ln b}&=&(2-\pi\rho^{'})g_{2,0} \nonumber \\
\frac{dg_{0,2}}{d\ln b}&=&(2-\pi\kappa^{'})g_{0,2} \nonumber \\
\frac{dg_{1,1}}{d\ln b}&=&\left(2-\frac{\pi}{4}(\rho^{'}+\kappa^{'})\right)g_{1,1} \nonumber \\
\frac{dg_{4}}{d\ln b}&=&(2-\frac{4}{\pi\kappa^{'}})g_{4} \nonumber \\
\frac{d\rho^{'}}{d\ln b}&=&-16g_{2,0}^{2}\rho^{'3}-\frac{g_{1,1}^{2}}{2}\rho^{'2}(\rho^{'}+\kappa^{'}) \nonumber \\
\frac{d\kappa^{'}}{d\ln b}&=&\frac{256g_{4}^{2}}{\pi^4\kappa^{'}}-16g_{0,2}^{2}\kappa^{'3}-\frac{g_{1,1}^{2}}{2}\kappa^{'2}(\rho^{'}+\kappa^{'}),
\end{eqnarray}

\begin{table}[!h]
\label{tab:1}
\centering
\caption{Fixed points of the coupling parameters under RG, and the corresponding phases.}\label{tab:1}
\begin{tabular}{|c|c|c|c|c|c|c|}
  \hline\hline
  $g_{2,0}$ & $g_{0,2}$ & $g_{4}$ & $g_{1,1}$ & $\rho^{'}$ & $\kappa^{'}$ & phase \\
  \hline
  $\infty$ & $\infty$ & 0 & $\infty$ & $0$ & $0$ & normal \\
  \hline
  0 & $\infty$ & 0 & 0 & $>8/\pi$ & $0$ & charge 4e SC \\
  \hline
  0 & 0 & $\infty$ & 0 & $>2/\pi$ & $\infty$ & chiral SC \\
  \hline
  $\infty$ & 0 & $\infty$ & 0 & $0$ & $\infty$ & chiral metal \\
  \hline\hline
\end{tabular}
\end{table}

The fixed points of N general RG flow equation $\frac{d\mathbf{g}}{d\ell}=R{(\mathbf{g})}$ is obtained by $R(\mathbf{g}^{\ast})=0$. The $\beta$ function of coupling constant which is very close to the fixed point $\mathbf{g}^{\ast}$ can be replaced by a linear mapping:
\begin{equation}
R(\mathbf{g})=R\left((\mathbf{g}-\mathbf{g}^{\ast})+\mathbf{g}^{\ast}\right)\simeq M(\mathbf{g}-\mathbf{g}^{\ast})
\end{equation}
where we have used $R(\mathbf{g}^{\ast})=0$, and $M_{\alpha\beta}=\frac{\partial R_{\alpha}}{\partial g_{\beta}}|_{\mathbf{g}=\mathbf{g}^{\ast}}$. To get the stability properties of the flow, we have to diagonalize the matrix $M_{N\times N}$. The eigenvalues denoted by $\lambda_{\alpha}$, $\alpha=1,2,...,N$. If the real parts of all the eigenvalues are negative or, at worst, zero, i.e. the scaling fields are all irrelevant or marginal. There are stable fixed points corresponding to the "stable phases". Complementary to the stable fixed points, if all the eigenvalues are positive and the scaling fields are all relevant, there are unstable fixed points. Additionally, there is generic class of fixed point with both relevant and irrelevant scaling fields. These points are associated with the boundary of the phase transition.

In the main text, although there are five possible RG flow results for the normal state in the first Table, only the first one actually appears in our calculations.
So in Table \ref{tab:1}., we present only four fixed points of the RG flow equation (\ref{eqn:RG-equation}) and the corresponding phases. They are consistent with the analysis and results shown in the text. Now, we analyze the stability of the four phases in the phase diagram.

(i)\textbf{The normal phase}:\\
We define $\bar{g}_{2,0}=\frac{1}{g_{2,0}}$, $\bar{g}_{0,2}=\frac{1}{g_{0,2}}$, $\bar{g}_{11}=\frac{1}{g_{11}}$ and $\ell=\ln b$ to simplify the calculation in the following. The RG flow equation can be rewritten as:
\begin{eqnarray}
\label{eqn:RG-equation-normal}
\frac{d\bar{g}_{2,0}}{d\ell}&=&-(2-\pi\rho^{'})\bar{g}_{2,0} \nonumber \\
\frac{d\bar{g}_{0,2}}{d\ell}&=&-(2-\pi\kappa^{'})\bar{g}_{0,2} \nonumber \\
\frac{d\bar{g}_{1,1}}{d\ell}&=&-\left(2-\frac{\pi}{4}(\rho^{'}+\kappa^{'})\right)\bar{g}_{1,1} \nonumber \\
\frac{dg_{4}}{d\ell}&=&(2-\frac{4}{\pi\kappa^{'}})g_{4} \nonumber \\
\frac{d\rho^{'}}{d\ell}&=&-16\bar{g}_{2,0}^{-2}\rho^{'3}-\frac{1}{2}\bar{g}_{1,1}^{-2}\rho^{'2}(\rho^{'}+\kappa^{'}) \nonumber \\
\frac{d\kappa^{'}}{d\ell}&=&\frac{256g_{4}^{2}}{\pi^4\kappa^{'}}-16\bar{g}_{0,2}^{-2}\kappa^{'3}-\frac{1}{2}\bar{g}_{1,1}^{-2}\kappa^{'2}(\rho^{'}+\kappa^{'}),
\end{eqnarray}
The $M$ matrix can be obtained as:
\begin{equation}
M=\left(
    \begin{array}{cccccc}
      -(2-\pi\rho^{'}) & 0 & 0 & 0 & \pi\bar{g}_{2,0} & 0 \\
      0 & -(2-\pi\kappa^{'}) & 0 & 0 & 0 & \pi\bar{g}_{0,2} \\
      0 & 0 & -\left(2-\frac{\pi}{4}(\rho^{'}+\kappa^{'})\right) & 0 & \frac{\pi}{4}\bar{g}_{1,1} &  \frac{\pi}{4}\bar{g}_{1,1}\\
      0 & 0 & 0 & 2-\frac{4}{\pi\kappa^{'}} & 0 & \frac{4g_{4}}{\pi\kappa^{'2}} \\
      \frac{32\rho^{'3}}{\bar{g}_{2,0}^{3}} & 0 & \frac{\rho^{'2}(\rho^{'}+\kappa^{'})}{\bar{g}_{1,1}^{3}} & 0 & -\frac{48\rho^{'2}}{\bar{g}_{2,0}^{2}}-\frac{3\rho^{'2}+2\rho^{'}\kappa^{'}}{2\bar{g}_{1,1}^{2}} & -\frac{\rho^{'2}}{2\bar{g}_{1,1}^{2}} \\
      0 & \frac{32\kappa^{'3}}{\bar{g}_{0,2}^{3}} & \frac{\kappa^{'2}(\rho^{'}+\kappa^{'})}{\bar{g}_{1,1}^{3}} & \frac{512g_{4}}{\pi^{4}\kappa^{'}} & -\frac{\kappa^{'2}}{2\bar{g}_{1,1}^{2}} & -\frac{256g_{4}^{2}}{\pi^{4}\kappa^{'2}}-\frac{48\kappa^{'2}}{\bar{g}_{0,2}^{2}}-\frac{2\kappa^{'}\rho^{'}+3\kappa^{'2}}{2\bar{g}_{1,1}^{2}} \\
    \end{array}
  \right)
\end{equation}
Now, let's analyze the order of four coupling parameters at the fixed point. We immediately have $\bar{g}_{0,2}\sim e^{-2\ell}$, $\bar{g}_{2,0}\sim e^{-2\ell}$, $\bar{g}_{1,1}\sim e^{-2\ell}$, and $g_{4}\sim e^{-\infty\ell}=0$. Then, we start to analyze the order of $\rho^{'}$ and $\kappa^{'}$. At the beginning, we can neglect the first term in the RG flow equation of stiffness $\kappa^{'}$ since the order of $g_{4}$ is extra low. We should discuss in three cases:\\
(a)If $O(\rho^{'})<O(\kappa^{'})$, we can only keep the highest order in the RG equation.
\begin{equation}
\frac{d\kappa^{'}}{d\ell}=-\frac{16\kappa^{'3}}{\bar{g}_{0,2}^{2}}-\frac{\kappa^{'3}}{2\bar{g}_{1,1}^{2}}\sim-e^{4\ell}\kappa^{'3}
\end{equation}
We immediately get the order of $\kappa^{'}$ as $e^{-2\ell}$. The differential equation of $\rho^{'}$ can be simplified as:
\begin{equation}
\frac{d\rho^{'}}{d\ell}=-\frac{\rho^{'2}\kappa^{'}}{2\bar{g}_{1,1}^{2}}
\end{equation}
The order of $\rho^{'}$ is $e^{-2\ell}$. So, we reach the result $O(\rho^{'})=O(\kappa^{'})\sim e^{-2\ell}$, which contradicts the previous assumption.\\
(b)If $O(\rho^{'})>O(\kappa^{'})$, the RG flow equation is simplified as:
\begin{eqnarray}
\frac{d\rho^{'}}{d\ell} \sim -e^{4\ell}\rho^{'3};\nonumber \\
\frac{d\kappa^{'}}{d\ell} \sim -\frac{\kappa^{'2}\rho^{'}}{2\bar{g}_{1,1}^{2}}
\end{eqnarray}
Solving the differential equations, we still have $O(\rho^{'})=O(\kappa^{'})\sim e^{-2\ell}$ which contradicts the previous assumption. \\
(c)If $O(\rho^{'})=O(\kappa^{'})$, we can arrive at the result $O(\rho^{'})=O(\kappa^{'})\sim e^{-2\ell}$ by the same method above. If we assume $\rho^{'}=\alpha e^{-2\ell}$ and $\kappa^{'}=\beta e^{-2\ell}$. Solving $\alpha$ and $\beta$, we have $\alpha=\beta=\pm \frac{2}{\sqrt{34}}$ or $\alpha=-\beta=\pm\sqrt{\frac{1}{8}}$.

Substituting the limit value of all the coupling constant and stiffness in the normal phase fixed point into the matrix $M$, we have:
\begin{equation}
M=\left(
    \begin{array}{cccccc}
      -2 & 0 & 0 & 0 & 0 & 0 \\
      0 & -2 & 0 & 0 & 0 & 0 \\
      0 & 0 & -2 & 0 & 0 & 0 \\
      0 & 0 & 0 & -\infty & 0 & 0 \\
      32\alpha^{3} & 0 & \alpha^{2}(\alpha+\beta) & 0 & -\frac{99\alpha^{2}+2\alpha\beta}{2} & -\frac{\alpha^{2}}{2} \\
      0 & 32\beta^{3} & \beta^{2}(\alpha+\beta) & 0 & -\frac{\beta^{2}}{2} & -\frac{99\beta^{2}+2\alpha\beta}{2} \\
    \end{array}
  \right)
\end{equation}
Obviously, all the eigenvalues of $M$ are negative which means that the normal phase is a stable fixed point.

(ii)\textbf{The charge $4e$ SC phase}:\\
By the same method above, we rewrite the RG flow equation as following to simplify the calculation:
\begin{eqnarray}
\label{eqn:RG-equation-4eSC}
\frac{dg_{2,0}}{d\ell}&=&(2-\pi\rho^{'})g_{2,0} \nonumber \\
\frac{d\bar{g}_{0,2}}{d\ell}&=&-(2-\pi\kappa^{'})\bar{g}_{0,2} \nonumber \\
\frac{dg_{1,1}}{d\ell}&=&\left(2-\frac{\pi}{4}(\rho^{'}+\kappa^{'})\right)g_{1,1} \nonumber \\
\frac{dg_{4}}{d\ell}&=&(2-\frac{4}{\pi\kappa^{'}})g_{4} \nonumber \\
\frac{d\rho^{'}}{d\ell}&=&-16g_{2,0}^{2}\rho^{'3}-\frac{1}{2}g_{1,1}^{2}\rho^{'2}(\rho^{'}+\kappa^{'}) \nonumber \\
\frac{d\kappa^{'}}{d\ell}&=&\frac{256g_{4}^{2}}{\pi^4\kappa^{'}}-16\bar{g}_{0,2}^{-2}\kappa^{'3}-\frac{1}{2}g_{1,1}^{2}\kappa^{'2}(\rho^{'}+\kappa^{'}),
\end{eqnarray}
We analyze the order of the coupling parameters. Firstly: $O(g_{4})\sim e^{-\infty\ell}$, $O(g_{2,0})<O(e^{-6\ell})$, $O(g_{0,2})\sim e^{2\ell}$, and $O(g_{1,1})<O(e^{0\ell})$. we keep the highest order term in the RG flow equation of $\kappa^{'}$:
\begin{equation}
\frac{d\kappa^{'}}{d\ell}=-16g_{0,2}^{2}\kappa^{'3}
\end{equation}
We obtain that $\kappa^{'}\sim e^{-2\ell}$. The differential matrix can be written as:
\begin{equation}
M=\left(
    \begin{array}{cccccc}
      2-\pi\rho^{'} & 0 & 0 & 0 & -\pi g_{2,0} & 0 \\
      0 & -(2-\pi\kappa^{'}) & 0 & 0 & 0 & \pi \bar{g}_{0,2} \\
      0 & 0 & 2-\frac{\pi}{4}(\rho^{'}+\kappa^{'}) & 0 & -\frac{\pi}{4}g_{1,1} &  -\frac{\pi}{4}g_{1,1}\\
      0 & 0 & 0 & 2-\frac{4}{\pi\kappa^{'}} & 0 & \frac{4g_{4}}{\pi\kappa^{'2}} \\
      -32\rho^{'3}g_{2,0} & 0 & -g_{1,1}\rho^{'2}(\rho^{'}+\kappa^{'}) & 0 & -48\rho^{'2}g_{2,0}^{2}-\frac{g_{1,1}^{2}(3\rho^{'2}+2\rho^{'}\kappa^{'})}{2} & -\frac{g_{1,1}^{2}\rho^{'2}}{2} \\
      0 & \frac{32\kappa^{'3}}{\bar{g}_{0,2}^{3}} & -g_{1,1}\kappa^{'2}(\rho^{'}+\kappa^{'}) & \frac{512g_{4}}{\pi^{4}\kappa^{'}} & -\frac{g_{1,1}^{2}\kappa^{'2}}{2} & -\frac{256g_{4}^{2}}{\pi\kappa^{'2}}-\frac{48\kappa^{'2}}{\bar{g}_{0,2}^{2}}-\frac{g_{1,1}^{2}(2\kappa^{'}\rho^{'}+3\kappa^{'2})}{2} \\
    \end{array}
  \right)
\end{equation}
Substituting all the orders of couplings and stiffness into $M$ matrix, we have:
\begin{equation}
M=\left(
    \begin{array}{cccccc}
      <-6 & 0 & 0 & 0 & 0 & 0 \\
      0 & -2 & 0 & 0 & 0 & 0 \\
      0 & 0 & <0 & 0 & 0 & 0 \\
      0 & 0 & 0 & -\infty & 0 & 0 \\
      0 & 0 & 0 & 0 & 0 & 0 \\
      0 & O(e^{0\ell}) & 0 & 0 & 0 & -O(e^{0\ell}) \\
    \end{array}
  \right)
\end{equation}
As we can see, all the eigenvalues are negative except the fifth one which is zero. Obviously, the charge $4e$ SC is a stable phase.

(iii)\textbf{The chiral SC phase}:\\
At the beginning, we rewrite the form of the RG flow equation:
\begin{eqnarray}
\label{eqn:RG-equation-chialSC}
\frac{dg_{2,0}}{d\ell}&=&(2-\pi\rho^{'})g_{2,0} \nonumber \\
\frac{dg_{0,2}}{d\ell}&=&(2-\pi\kappa^{'})g_{0,2} \nonumber \\
\frac{dg_{1,1}}{d\ell}&=&\left(2-\frac{\pi}{4}(\rho^{'}+\kappa^{'})\right)g_{1,1} \nonumber \\
\frac{d\bar{g}_{4}}{d\ell}&=&-(2-\frac{4}{\pi\kappa^{'}})\bar{g}_{4} \nonumber \\
\frac{d\rho^{'}}{d\ell}&=&-16g_{2,0}^{2}\rho^{'3}-\frac{1}{2}g_{1,1}^{2}\rho^{'2}(\rho^{'}+\kappa^{'}) \nonumber \\
\frac{d\kappa^{'}}{d\ell}&=&\frac{256}{\pi^4\kappa^{'}\bar{g}_{4}^{2}}-16g_{0,2}^{2}\kappa^{'3}-\frac{1}{2}g_{1,1}^{2}\kappa^{'2}(\rho^{'}+\kappa^{'}),
\end{eqnarray}
We analyze the order of all the coupling constants and the stiffness parameters. $O(g_{2,0})<e^{0\ell}$, $O(g_{0,2})\sim e^{-\infty\ell}$, $O(g_{1,1})\sim e^{-\infty\ell}$, and $O(g_{4})\sim e^{2\ell}$. We don't need analyze the order of $\rho^{'}$, since it is a fixed parameter in the region $>\frac{2}{\pi}$. And in the last RG flow equation, $g_{4}^{2}$ has the highest order obviously. So, we can neglect other terms in that equation and solve the differential equation to get the order of $\kappa^{'}$. $\kappa^{'}\sim e^{2\ell}$ is obtained.

The differential matrix can be written as:
\begin{equation}
M=\left(
    \begin{array}{cccccc}
      2-\pi\rho^{'} & 0 & 0 & 0 & -\pi g_{2,0} & 0 \\
      0 & (2-\pi\kappa^{'}) & 0 & 0 & 0 & -\pi g_{0,2} \\
      0 & 0 & 2-\frac{\pi}{4}(\rho^{'}+\kappa^{'}) & 0 & -\frac{\pi}{4}g_{1,1} &  -\frac{\pi}{4}g_{1,1}\\
      0 & 0 & 0 & -(2-\frac{4}{\pi\kappa^{'}}) & 0 & -\frac{4\bar{g}_{4}}{\pi\kappa^{'2}} \\
      -32\rho^{'3}g_{2,0} & 0 & -\frac{\rho^{'2}(\rho^{'}+\kappa^{'})}{\bar{g}_{1,1}} & 0 & -48\rho^{'2}g_{2,0}^{2}-\frac{g_{1,1}^{2}(3\rho^{'2}+2\rho^{'}\kappa^{'})}{2} & -\frac{g_{1,1}^{2}\rho^{'2}}{2} \\
      0 & -32\kappa^{'3}g_{0,2} & -\frac{\kappa^{'2}(\rho^{'}+\kappa^{'})}{\bar{g}_{1,1}} & -\frac{512}{\pi^{4}\kappa^{'}\bar{g}_{4}^{3}} & -\frac{g_{1,1}^{2}\kappa^{'2}}{2} & -\frac{256g_{4}^{2}}{\pi\kappa^{'2}}-\frac{48\kappa^{'2}}{\bar{g}_{0,2}^{2}}-\frac{g_{1,1}^{2}(2\kappa^{'}\rho^{'}+3\kappa^{'2})}{2} \\
    \end{array}
  \right)
\end{equation}
Substituting all the limit values of the coupling constant and stiffness parameters at the fixed point, we have:
\begin{equation}
M=\left(
    \begin{array}{cccccc}
      <0 & 0 & 0 & 0 & 0 & 0 \\
      0 & -\infty & 0 & 0 & 0 & 0 \\
      0 & 0 & -\infty & 0 & 0 & 0 \\
      0 & 0 & 0 & -2 & 0 & 0 \\
      0 & 0 & 0 & 0 & 0 & 0 \\
      0 & 0 & 0 & -\infty & 0 & -O(e^{0\ell}) \\
    \end{array}
  \right)
\end{equation}
As we can see, all the eigenvalues are negative except the fifth one which is zero. Obviously, the chiral SC is a stable phase.

(iv)\textbf{The chiral metal phase}:\\
We rewrite the RG flow equation:
\begin{eqnarray}
\label{eqn:RG-equation-chiralmetal}
\frac{d\bar{g}_{2,0}}{d\ell}&=&-(2-\pi\rho^{'})\bar{g}_{2,0} \nonumber \\
\frac{dg_{0,2}}{d\ell}&=&(2-\pi\kappa^{'})g_{0,2} \nonumber \\
\frac{dg_{1,1}}{d\ell}&=&\left(2-\frac{\pi}{4}(\rho^{'}+\kappa^{'})\right)g_{1,1} \nonumber \\
\frac{d\bar{g}_{4}}{d\ell}&=&-(2-\frac{4}{\pi\kappa^{'}})\bar{g}_{4} \nonumber \\
\frac{d\rho^{'}}{d\ell}&=&-16\bar{g}_{2,0}^{-2}\rho^{'3}-\frac{g_{1,1}^{2}}{2}\rho^{'2}(\rho^{'}+\kappa^{'}) \nonumber \\
\frac{d\kappa^{'}}{d\ell}&=&\frac{256\bar{g}_{4}^{-2}}{\pi^4\kappa^{'}}-16g_{0,2}^{2}\kappa^{'3}-\frac{g_{1,1}^{2}}{2}\kappa^{'2}(\rho^{'}+\kappa^{'}),
\end{eqnarray}
The order of each constant can be analyzed by the same method as before. $g_{2,0}\sim e^{2\ell}$, $g_{0,2}\sim e^{-\infty\ell}$, $g_{1,1}\sim e^{-\infty\ell}$, $g_{4}\sim e^{2\ell}$, and the RG flow equation of stiffness can be simplified as:
\begin{eqnarray}
\frac{d\rho^{'}}{d\ell}=-\frac{16\rho^{'3}}{\bar{g}_{2,0}^{2}};\nonumber \\
\frac{d\kappa^{'}}{d\ell}=\frac{256}{\pi^{4}\kappa^{'}\bar{g}_{4}^{2}}
\end{eqnarray}
Solving the differential equation above, we have $\rho^{'}\sim e^{-2\ell}$ and $\kappa^{'}\sim e^{2\ell}$. At the same time, we can write out the $M$ matrix as following:
\begin{equation}
M=\left(
    \begin{array}{cccccc}
      -(2-\pi\rho^{'}) & 0 & 0 & 0 & \pi\bar{g}_{2,0} & 0 \\
      0 & 2-\pi\kappa^{'} & 0 & 0 & 0 & -\pi g_{0,2} \\
      0 & 0 & 2-\frac{\pi}{4}(\rho^{'}+\kappa^{'}) & 0 & -\frac{\pi}{4}g_{1,1} &  -\frac{\pi}{4}g_{1,1}\\
      0 & 0 & 0 & -(2-\frac{4}{\pi\kappa^{'}}) & 0 & -\frac{4\bar{g}_{4}}{\pi\kappa^{'2}} \\
      \frac{32\rho^{'3}}{\bar{g}_{2,0}^{3}} & 0 & -\frac{\rho^{'2}(\rho^{'}+\kappa^{'})}{\bar{g}_{1,1}} & 0 & -\frac{48\rho^{'2}}{\bar{g}_{2,0}^{2}}-\frac{g_{1,1}^{2}(3\rho^{'3}+2\rho^{'}\kappa^{'})}{2} & -\frac{g_{1,1}^{2}\rho^{'2}}{2} \\
      0 & -32\kappa^{'3}g_{0,2} & -\frac{\kappa^{'2}(\rho^{'}+\kappa^{'})}{\bar{g}_{1,1}} & -\frac{512}{\pi^{4}\kappa^{'}\bar{g}_{4}^{3}} & -\frac{g_{1,1}^{2}\kappa^{'2}}{2} & -\frac{256\bar{g}_{4}^{-2}}{\pi^{4}\kappa^{'2}}-48\kappa^{'2}g_{0,2}^{2}-\frac{(2\kappa^{'}\rho^{'}+3\kappa^{'2})}{2\bar{g}_{1,1}^{2}} \\
    \end{array}
  \right)
\end{equation}
Substituting all the values of the coupling constant and stiffness parameters in the fixed point, we arrive at:
\begin{equation}
M=\left(
    \begin{array}{cccccc}
      -2 & 0 & 0 & 0 & 0 & 0 \\
      0 & -\infty & 0 & 0 & 0 & 0 \\
      0 & 0 & -\infty & 0 & 0 & 0 \\
      0 & 0 & 0 & -2 & 0 & 0 \\
      O(e^{0\ell}) & 0 & 0 & 0 & -O(e^{0\ell}) & 0 \\
      0 & 0 & 0 & -\infty & 0 & -O(e^{0\ell}) \\
    \end{array}
  \right)
\end{equation}
All the eigenvalues are negative. Obviously, the chiral metal is also a stable phase.

\section{More detailed Results about the RG study}

To compare with the phase diagrams with different initial value of the coupling parameters, we present Fig.(\ref{phases_RG}) in this section. As shown in this figure, we find the direct transition regime between chiral TSF and normal phase are enhanced with larger initial value of half-vortices couplings $g_{1,1}$. Additionally, chiral metal phase will be enlarged in the phase diagram if we increase the initial value of the coupling parameter $g_{4}$.

Our RG results indicate that the interesting phases of charge 4e SC and chiral metal can always exist with different initial coupling parameters.

\begin{figure}[h]
	\centering
	\includegraphics[width=1.0\textwidth]{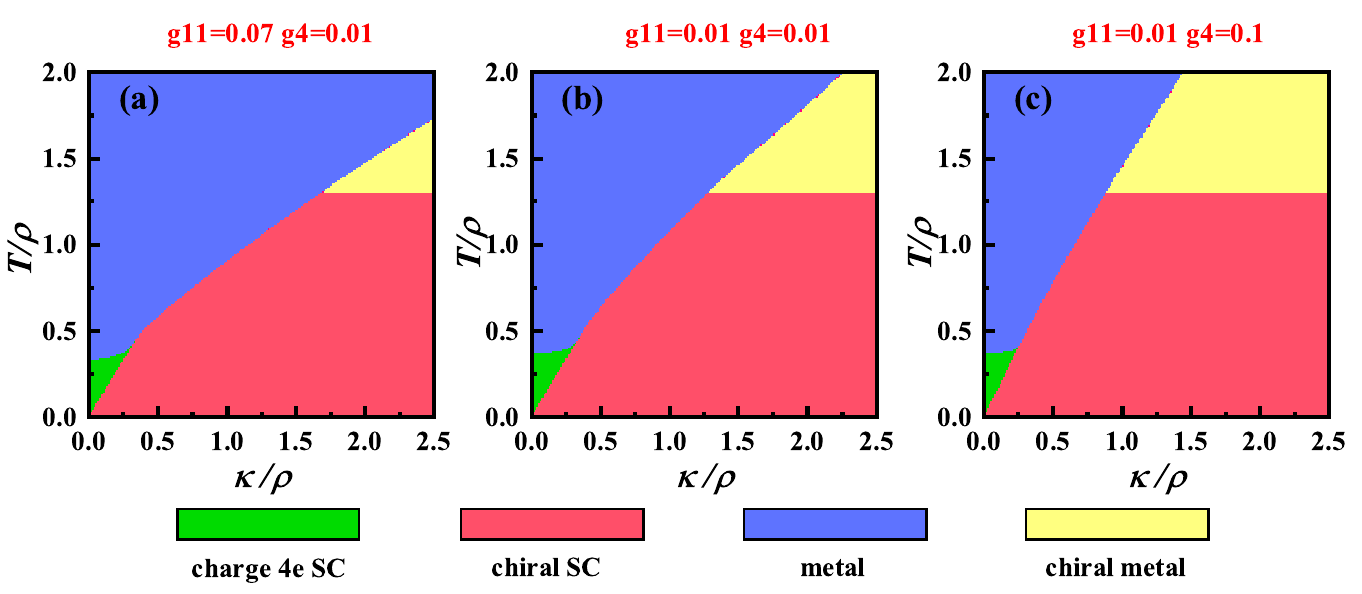}
	\caption{\label{phases_RG}(Color online) Phase diagram provided by the RG approach with different initial coupling parameters. The initial values of the coupling parameters are $g_{2,0}=g_{0,2}=0.1$, $g_{1,1}=0.07$ and $g_{4}=0.01$ for (a), $g_{2,0}=g_{0,2}=0.1$, $g_{1,1}=g_{4}=0.01$ for (b), and $g_{2,0}=g_{0,2}=0.1$, $g_{1,1}=0.01$, $g_{4}=0.1$ for (c).}
\end{figure}

\section{More details Results about the MC study}

\begin{figure}[h]
	\centering
	\includegraphics[width=0.75\textwidth]{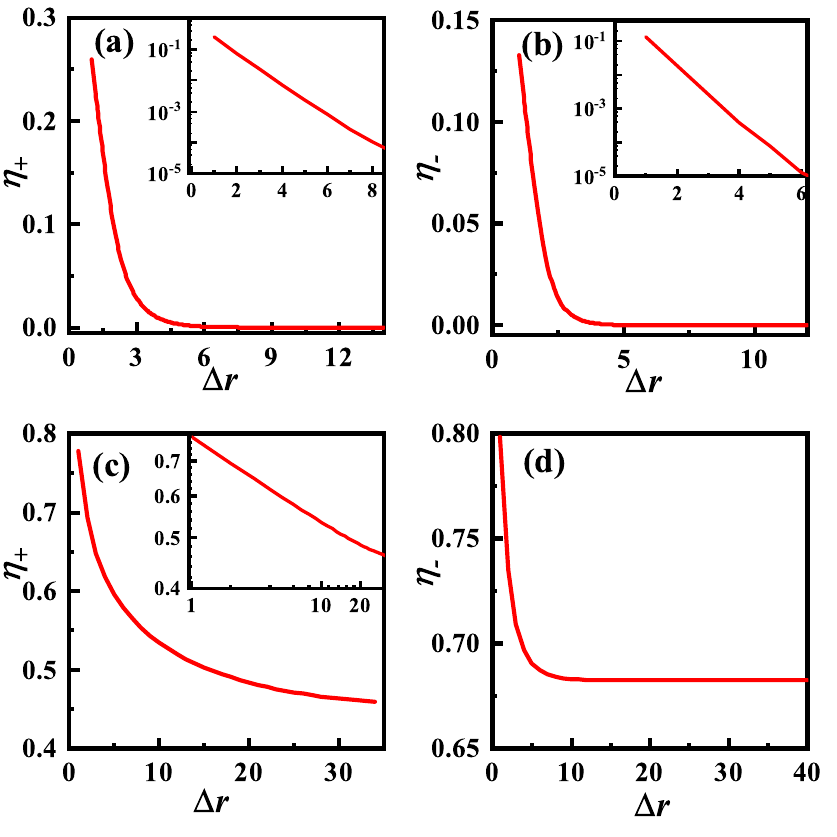}
	\caption{(Color online) (a-b) The correlation function $\eta_{\pm}$ for the parameter point B($\kappa/\rho=0.6, T/\rho=0.45$) in Fig.2(b) in the main text, respectively. The y- axes of the inset are logarithmic axes. (c) The correlation function $\eta_{+}$ for the parameter point C($\kappa/\rho=1, T/\rho=0.2$) in Fig.2(b) in the main text, both the x- and y- axes of the inset are logarithmic axes. (d) The correlation function $\eta_{-}$ for the parameter point C in Fig.2(b) in the main text, the y- axis of the inset is logarithmic axis.}\label{BC}
\end{figure}

\begin{figure}[htbp]
	\centering
	\includegraphics[width=0.75\textwidth]{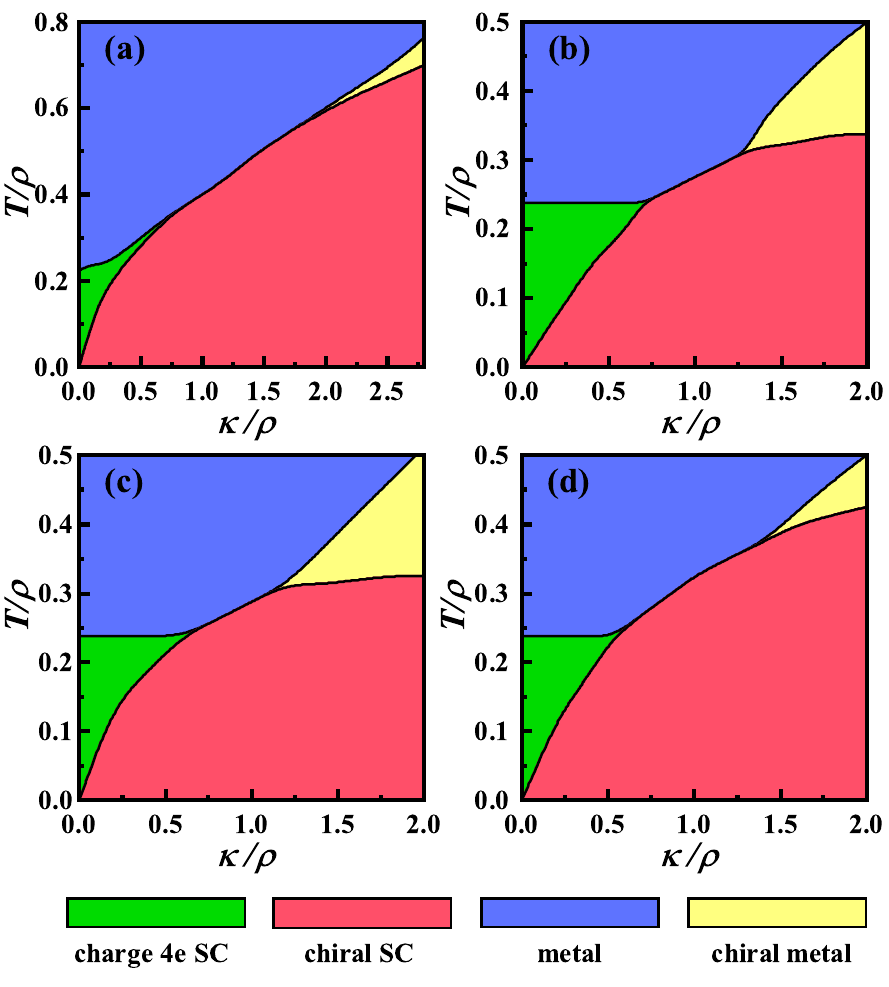}
	\caption{(Color online) Phase diagram provided by the MC study with different parameter $\gamma$ and extra $B$ term. (a) The same parameters as those in the phase diagram in the main text except that $\gamma=\rho\kappa/2(\rho+\kappa)$. (b) The same parameters as those in the phase diagram in the main text except that $\gamma=\rho\kappa/6(\rho+\kappa)$. (c) The same parameters as those in the phase diagram in the main text except that $\gamma=0.1\rho$ is a constant. (d)The same parameters as those in the phase diagram in the main text except that a weak first-order Josephson coupling with coefficient $B=0.01\rho$ is added.}\label{phases}
\end{figure}

\begin{figure}[htbp]
	\centering
	\includegraphics[width=0.95\textwidth]{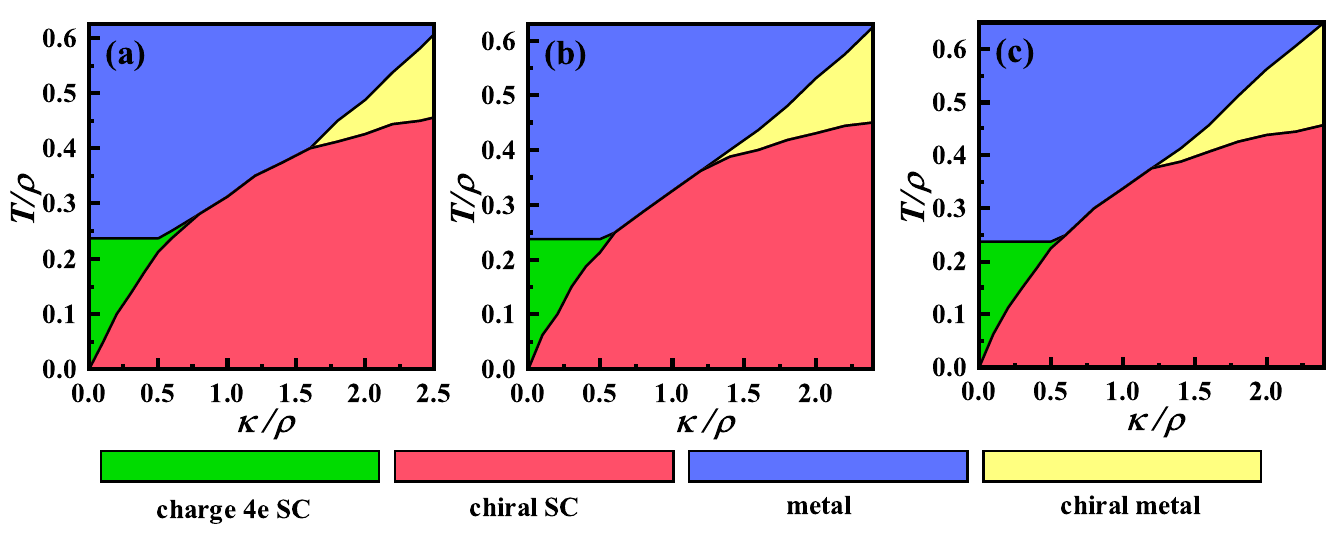}
	\caption{(Color online) Phase diagram provided by the MC study with different parameter $A$. (a-c) The same parameters as those in the phase diagram in the main text except that $A=0.0125\rho,0.05\rho$ and $0.1\rho$, respectively.}\label{phases1}
\end{figure}

The properties of the correlation function of the parameter point B and C in phase diagram is shown in Fig.~\ref{BC}. For the parameter point B, Fig.~\ref{BC}(a) and (b) show the correlation functions $\eta_{+}$ and $\eta_{-}$, respectively. Both the correlation function $\eta_{+}$ and $\eta_{-}$ are exponentially decay, which proves that point parameter B is the metal state. On the contrary, For the parameter point C, Fig.~\ref{BC}(c) and (d) show the correlation functions $\eta_{+}$ and $\eta_{-}$, respectively. The correlation function $\eta_{+}$ is power law decay but the correlation function $\eta_{-}$ is a constant, which proves that parameter point D is the chiral SC.

To verify the generality of the discretized Hamiltonian, we perform the MC study with different $\gamma$ and $A$ to obtain the phase diagram, shown in Fig.~\ref{phases} (a-c) and Fig.~\ref{phases1} (a-c). The phase diagrams for different $\gamma$ do not change qualitatively.

To verify the stability of the results, we perform the MC study with a weak first-order Josephson-coupling term added, whose coefficient is $B=0.01\rho$. The $\gamma$ is the same as that adopted in the main text. The phase diagram is shown in Fig.~\ref{phases}(d), which is similar with that obtained for zero $B$.

	\section{The MC result without considering kinematic constraint($\gamma=0$).}
	In order to highlight the importance of $\gamma$ term in the Eq. (15) in main text, we calculate the phase diagram with $\gamma=0$ and present the theoretical explanation about this phase diagram.
	
	If we turn off the $\gamma$ term in Eq. (15) in main text, we have $\alpha=\rho/4, \lambda=\kappa/4$. Then we redo the Monte-Carlo calculations. Consequently, the obtained phase diagram is displayed in the following Fig.~\ref{phases3} (a). This phase diagram is very simple, which is divided by two lines into four phases touching at one qua-critical point. The straight line parallel to the x-axis represents the K-T transition, suggesting that the (quasi-) ordering temperature of $\theta_+$ only relies on $\rho$. The line passing through the coordinate origin represents the Ising transition, suggesting that the ordering temperature of $\theta_-$ only relies on $\kappa$ when fixing A. This phase diagram suggests that $\theta_+$ and $\theta_-$ are decoupled, which is analytically understood as follow.
	
	\begin{figure}[htbp]
		\centering
		\includegraphics[width=0.75\textwidth]{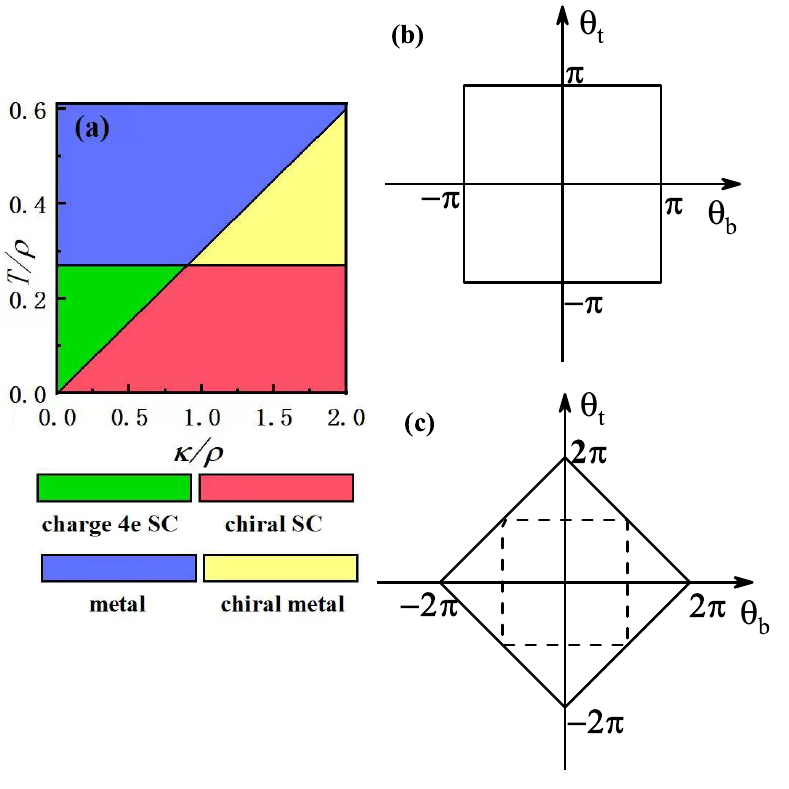}
		\caption{(Color online) (a) Phase diagram provided by the MC study with different parameter $\gamma$. The same parameters as those in the phase diagram in the main text except that $\gamma=0$. (b) Integral region of $\theta_b$ and $\theta_t$ at a given site. (c) Expanded integral region of $\theta_b$ and $\theta_t$ at a given site.}\label{phases3}
	\end{figure}
	
	The partition function of the model is written as
	\begin{eqnarray}\label{z1}	
		Z=\int\dots\int\prod_{\vec{r}_i}d\theta_{t}(\vec{r}_i)d\theta_{b}(\vec{r}_i)e^{-\beta H[\{\theta_{t}(\vec{r}_i),\theta_{b}(\vec{r}_i)\}]}
	\end{eqnarray}
	For each site, the integral region is within the ``first Brilloiun Zone (BZ)'' $\theta_t \in [-\pi,\pi), ~ \theta_b \in [-\pi,\pi)$ shown in Fig.~\ref{phases3} (b). Since $H[\{\theta_{t}(\vec{r}_i),\theta_{b}(\vec{r}_i)\}]$ is a periodic function of $\theta_b$ and $\theta_t$ with period $2\pi$, the integral region can be expanded to the ``second BZ'' shown in Fig.~\ref{phases3} (c), i.e. $\theta_t+\theta_b \in [-2\pi,2\pi), ~ \theta_t-\theta_b \in [-2\pi,2\pi)$ or equivalently $\theta_+ \in [-\pi,\pi), ~ \theta_- \in [-\pi,\pi)$. Such an expansion only doubles Z, and would not change the physics. For $\gamma=0$, we have $H=H_+[\{\theta_+(\vec{r})\}]+H_-[\{\theta_-(\vec{r})\}]$, and then we have
	
	\begin{eqnarray}\label{z2}	
		Z &=& \int\cdot\cdot\cdot\int\prod_{\vec{r}_i}d\theta_{+}(\vec{r}_i)d\theta_{-}(\vec{r}_i)e^{-\beta H_+[\{\theta_{+}(\vec{r}_i)\}]}\cdot e^{-\beta H_-[\{\theta_{-}(\vec{r}_i)\}]}\nonumber\\
		&=& \int\dots\int\prod_{\vec{r}_i}d\theta_{+}(\vec{r}_i)e^{-\beta H_+[\{\theta_{+}(\vec{r}_i)\}]}\cdot \int\dots\int\Pi_{\vec{r}_i}d\theta_{-}(\vec{r}_i)e^{-\beta H_-[\{\theta_{-}(\vec{r}_i)\}]}\nonumber\\
		&=& Z_+\cdot Z_-
	\end{eqnarray}
	This result explains why $\theta_+$ and $\theta_-$ are decoupled for $\gamma=0$.
	
	However, since we do not consider kinematically correlated of $\theta_+$ and $\theta_-$, the phase diagram shown in the Fig.~\ref{phases3} (a) is topologically different from Fig. 2 in the main text and is wrong.

\end{widetext}
\end{document}